\documentclass[12pt,oneside]{article}
\usepackage{amssymb,amsmath,latexsym,amsthm,amsfonts}
\usepackage[english]{babel}
\usepackage[T1]{fontenc}
\usepackage{color,hyperref}
\numberwithin{ex}{section} \numberwithin{rem}{section}
\numberwithin{equation}{section} \numberwithin{thm}{section}
\numberwithin{lem}{section} \numberwithin{coro}{section}

\def\1g{1\hskip -3pt \mbox{l}}

\small\normalsize

\title{Autoregressive order identification for VAR models with non-constant variance}

\author{
{\sc Hamdi Ra\"{i}ssi \footnote{IRMAR-INSA, 20 avenue des buttes de
Coësmes, CS 70839, F-35708 Rennes Cedex 7, France. }}}

\begin{document}

\maketitle  \noindent {\em Abstract:}\quad The identification of the
lag length for vector autoregressive models by mean of Akaike
Information Criterion ($AIC$), Partial Autoregressive and
Correlation Matrices (PAM and PCM hereafter) is studied in the
framework of processes with time varying variance. It is highlighted
that the use of the standard tools are not justified in such a case.
As a consequence we propose an adaptive $AIC$ which is robust to the
presence of unconditional heteroscedasticity.
Corrected confidence bounds are proposed for the usual PAM and PCM
obtained from the Ordinary Least Squares (OLS) estimation. We also use the
unconditional variance structure of the innovations to develop adaptive
PAM and PCM. It is underlined that the adaptive PAM and PCM are more
accurate than the OLS PAM and PCM for identifying the lag length of
the autoregressive models. Monte Carlo experiments show that the
adaptive $AIC$ have a greater ability to select the correct
autoregressive order than the standard $AIC$. An illustrative
application using US international finance data is presented.

\vspace*{.7cm} \noindent {\em Keywords:} Time varying unconditional variance;
VAR model; Model selection; AIC; Partial autoregressive matrices;
Partial autoregressive matrices.\\

\section{Introduction}

The analysis of time series using linear models is usually carried
out following three steps. First the model is identified, then
estimated and finally we proceed to the checking of the
goodness-of-fit of the model (see Brockwell and Davis (1991,
chapters 8 and 9)). Tools for the three phases in the
specification-estimation-verification modeling cycle of time series
with constant unconditional innovations variance are available in
any of the specialized softwares as for instance R, SAS or JMulTi.
The identification stage is important for the choice of a suitable
model for the data. In this step the partial autoregressive and
correlation matrices (PAM and PCM hereafter) are often used to
identify VAR models with stationary innovations (see Tiao and Box
(1981)). Information criteria are also extensively used. In the
framework of stationary processes numerous information criteria have
been studied (see e.g. Hannan and Quinn (1979), Cavanaugh (1997) or
Boubacar Mainassara (2012)). One of the most commonly used
information criterion is the Akaike Information Criterion ($AIC$)
proposed by Akaike (1973). Nevertheless it is widely documented in
the literature that the constant variance assumption is
unrealistic for many economic data. Reference can be made to
Mc-Connell, Mosser and Perez-Quiros (1999), Kim and Nelson (1999),
Stock and Watson (2002), Ahmed, Levin and Wilson (2002), Herrera and
Pesavento (2005) or Davis and Kahn (2008). In this paper we
investigate the lag length identification problem of autoregressive
processes in the important case where the unconditional innovations
variance is time varying.

The statistical inference of processes with non constant variance
has recently attracted much attention. Horv\'{a}th and Steinebach
(2000), Sanso, Arago and Carrion (2004) or Galeano and Pena (2007)
among other contributions proposed tests to detect variance and/or covariance breaks
in the residuals. Francq and Gautier (2004) studied the estimation
of ARMA models with time varying parameters, allowing a finite
number of regimes for the variance. Mikosch and St\u{a}ric\u{a}
(2004) give some theoretical evidence that financial data may
exihibit non constant variance. In the context of GARCH models
reference can be made to the works of Kokoszka and Leipus (2000),
Engle and Rangel (2008), Dahlhaus and Rao (2006) or Horvath,
Kokoszka and Zhang (2006) who investigated the inference for
processes with possibly unconditional time varying variance. In the
multivariate framework Bai (2000), Qu and Perron (2007) or Kim and
Park (2010)
among others studied autoregressive models with unconditionally non
constant variance. Aue, H\"{o}rmann, Horv\`{a}th and Reimherr (2009)
studied the break detection in the covariance structure of
multivariate processes.
Xu and Phillips (2008) studied the estimation of univariate
autoregressive models whose innovations have a non constant
variance. Patilea and Raïssi (2012) generalized their findings
in the case of Vector AutoRegressive (VAR) models with time-varying
variance
and found that the
asymptotic covariance matrix obtained if one take into account of
the non constant variance can be quite different from the standard
covariance matrix expression. As a consequence they also provided
Adaptive Least Squares (ALS) estimators which achieve a more
efficient estimation of the autoregressive parameters. Patilea and
Raïssi (2011) proposed tools for checking the adequacy of the
autoregressive order of VAR models when the unconditional variance
is non constant.

In this paper modified tools for lag length identification in the
case of multivariate autoregressive processes with time-varying variance are
introduced. The unreliability of the use of the standard $AIC$ for
the identification step in VAR modeling in presence of non constant
variance is first highlighted. Consequently a modified $AIC$ based
on the adaptive estimation of the non constant variance structure
is proposed. We establish the suitability of the adaptive $AIC$ to
identify the autoregressive order of non stationary but stable VAR
processes through theoretical results and numerical illustrations.
On the other hand it is also shown that the standard results on the
OLS estimators of the PAM and PCM can be quite misleading.
Consequently corrected confidence bounds are proposed. Using the
adaptive approach more efficient estimators of the PAM and PCM are
proposed. Therefore the identification tools proposed in this paper
may be viewed as a complement of the above mentioned results on the
estimation and diagnostic testing in the important framework of
autoregressive models with non constant variance.

The structure of the paper is as follow. In Section \ref{S2} we
define the model and introduce assumptions which give the general
framework of our study. The asymptotic behavior of different
estimators of the autoregressive parameters is given. We also
describe the adaptive estimation of the variance. In Section
\ref{S3} it is shown that the standard $AIC$ is irrelevant for model
selection when the innovations variance is not constant. The
adaptive $AIC$ is derived taking into account the time-varying
variance in the Kullback-Leibler discrepancy. In Section \ref{S4}
some Monte Carlo experiments results are given to examine the
performances of the studied information criteria for VAR model
identification in our non standard framework. We also investigate
the lag length selection of a bivariate system of US international
finance variables.

\section{Estimation of the model}
\label{S2}

In this paper we restrict our attention to VAR models since they are
extensively used for the analysis of multivariate time series (see
e.g. L\"utkepohl (2005)). Let us consider the $d$-dimensional
autoregressive process $(X_t)$ satisfying

\begin{eqnarray}\label{VAR}
&&X_t=A_{01}X_{t-1}+\dots+A_{0p}X_{t-p_0}+u_t\\&&
u_t=H_t\epsilon_t,\nonumber
\end{eqnarray}
where the $A_{0i}$'s, $i\in\{1,\dots,p_0\}$, are such that
$\det(A(z))\neq 0$ for all $|z|\leq 1$, with
$A(z)=1-\sum_{i=1}^{p_0}A_{0i} z^i$ and $\det(.)$ denotes the
determinant of a square matrix. We suppose that
$X_{-p+1},\dots,X_0,\dots,X_n$ are observed with $p>p_0$.
Now let us denote by $[.]$ the integer part. For ease of exposition
we shall assume that the process $(\epsilon_t)$ is iid multivariate standard
Gaussian. Throughout the paper we assume that
the following conditions on the unconditional variance structure of the innovations process $(u_t)$ hold.\\

\textbf{Assumption A1:} \quad \textit{The $d\times d$ matrices
$H_{t}$ are positive definite and satisfy $H_{[Tr]}=G(r)$, where the
components of the matrix $G(r):=\{g_{kl}(r)\}$ are measurable
deter\-ministic functions on the interval $(0,1]$, such that
$\sup_{r\in(0,1]}|g_{kl}(r)|<\infty$, and each $g_{kl}$ satisfies a
Lipschitz condition piecewise on a finite number of some
sub-intervals that partition $(0,1]$.
The matrix $\Sigma(r)=G(r)G(r)'$ is assumed positive definite for all $r$.}\\

The rescaling method of Dahlhaus (1997) is considered to specify the
unconditional variance structure in Assumption {\bf A1}. Note that one should
formally use the notation $X_{t,n}$ with $0<t\leq n$ and
$n\in\mathbb{N}$. Nevertheless we do not use the subscript $n$ to
lighten the notations. This specification allows to consider kinds of
time-varying variance which are commonly considered in the
literature as for instance abrupt shifts, smooth transitions or
periodic heteroscedasticity. Note that Sensier and Van Dijk (2004)
found that approximately 80\% among 214 US macro-economic data they
investigated exhibit a variance break. St\u{a}ric\u{a} (2003)
hypothesized that the returns of the Standard and Poors 500 stock
market index have a non constant unconditional variance. Then
considering the framework given by {\bf A1} is important given the
strong empirical evidence of non-constant unconditional variance
in many macro-economic and financial data. Our assumption is similar
to that of recent papers in the literature. For instance similar
structure for the variance was considered by Xu and Phillips
(2008) or Kim and Park (2010) among others. Our framework encompass
the important case of piecewise constant variance as considered in
Pesaran and Timmerman (2004) or Bai (2000).
Finally it is important to underline that the framework induced by {\bf A1} is different from the case of autoregressive processes with conditionally heteroscedastic but (strictly) stationary errors. For instance models like the GARCH or the All-Pass models cannot take into account for non constant unconditional variance in the innovations. The model identification problem for stationary processes which may display nonlinearities has been recently investigated by Boubacar Mainassara (2012) in a quite general framework.\\

In this part we introduce estimators of the autoregressive
parameters. Let us rewrite (\ref{VAR}) as follow

\begin{eqnarray}\label{VARvec}
&& X_t=(\tilde{X}_{t-1}'\otimes I_d)\theta_0+u_t\\&&
u_t=H_t\epsilon_t,\nonumber
\end{eqnarray}
where $\theta_0=(\mbox{vec}\:(A_{01})',\dots,\mbox{vec}\:(
A_{0p_0})')'\in\mathbb{R}^{p_0d^2}$ is the vector of the true
autoregressive parameters and
$\tilde{X}_{t-1}=(X_{t-1}',\dots,X_{t-p_0}')'$. For a fitted
autoregressive order $p\geq p_0$, the  OLS estimator is given by

\begin{equation*}
\hat{\theta}_{OLS}=\hat{\Sigma}_{\tilde{X}}^{-1}\mbox{vec}\:\left(\hat{\Sigma}_{X}\right),
\end{equation*}
where
$$\hat{\Sigma}_{\tilde{X}}=n^{-1}\sum_{t=1}^n\tilde{X}_{t-1}^p(\tilde{X}_{t-1}^{p})'\otimes
I_d\quad\mbox{and}\quad\hat{\Sigma}_X=n^{-1}
\sum_{t=1}^nX_t(\tilde{X}_{t-1}^{p})',$$ and
$\tilde{X}_{t-1}^p=(X_{t-1}',\dots,X_{t-p}')'$. If we suppose that
the true unconditional covariance matrices $\Sigma_t:=H_tH_t'$ are known, we
can define the following Generalized Least Squares (GLS) estimator

\begin{equation}\label{GLS}
\hat{\theta}_{GLS}=\hat{\Sigma}_{\tilde{\underline{X}}}
^{-1}\mbox{vec}\:\left(\hat{\Sigma}_{\underline{X}}\right),
\end{equation}
with
$$\hat{\Sigma}_{\tilde{\underline{X}}}=n^{-1}\sum_{t=1}^n\tilde{X}_{t-1}^{p}(\tilde{X}_{t-1}^{p})'
\otimes\Sigma_t^{-1}
\quad\mbox{and}\quad\hat{\Sigma}_{\underline{X}}=n^{-1}
\sum_{t=1}^n\Sigma_t^{-1}X_t(\tilde{X}_{t-1}^{p})'.$$

Let us define
$u_t(\theta)=X_t-\{(\tilde{X}_{t-1}^{p})'\otimes I_d\}\theta$ with
$\theta\in\mathbb{R}^{pd^2}$. Note that $\hat{\theta}_{GLS}$
maximizes the conditional log-likelihood function (up to a constant and divided
by $n$)

\begin{equation}\label{likeals}
\mathcal{L}_{GLS}(\theta)=-\frac{1}{2n}\sum_{t=1}^n\ln\left\{\det(\Sigma_t)\right\}-u_t(\theta)'\Sigma_t^{-1}u_t(\theta),
\end{equation}
(see L\"{u}tkepohl (2005, p 589)). If we assume that the innovations
process variance is constant ($\Sigma_t=\Sigma_{u}$ for all $t$)
and unknown, the standard conditional log-likelihood function

\begin{equation}\label{OLSlike}
\mathcal{L}_{OLS}(\theta,\Sigma)=-\frac{1}{2}\ln(\det(\Sigma))
-\frac{1}{2n}\sum_{t=1}^nu_t(\theta)'\Sigma^{-1}u_t(\theta)
\end{equation}
where $\Sigma$ is a $d\times d$ invertible matrix, is usually used
for the estimation of the parameters.
The estimator obtained by maximizing $\mathcal{L}_{OLS}$ with
respect to $\theta$ corresponds to $\hat{\theta}_{OLS}$.
In this case the estimator of the constant variance $\Sigma_{u}$ is given by $\hat{\Sigma}_u:=n^{-1}\sum_{t=1}^n\hat{u}_t\hat{u}_t'$ where $\hat{u}_t:=u_t(\hat{\theta})$ are the residuals of the OLS estimation of (\ref{VARvec}).\\

In practice the assumption of known variance is unrealistic.
Therefore we consider an adaptive estimator of the autoregressive
parameters. We may first define adaptive estimators of the true
unconditional variances $\Sigma_t:=H_tH_t'$ as in Patilea and Raïssi
(2012)


$$\check{\Sigma}_t=\sum_{i=1}^n w_{ti} (b)\hat{u}_i\hat{u}_i',$$
where the weights $w_{ti}$ are given by
$$w_{ti}(b)= \left(\sum_{i=1}^nK_{ti}(b)\right)^{-1} K_{ti}(b),$$
with $b$ the bandwidth and
$$K_{ti} (b) =\left\{
              \begin{array}{c}
                K(\frac{t-i}{nb})\quad \mbox{if}\quad t\neq i\\
                0  \quad\mbox{if}\quad t=i,\\
              \end{array}
            \right.$$
where $K(.)$ is the kernel function which is such that
$\int_{-\infty}^{+\infty} K(z) dz=1$. The bandwidth $b$ is taken in
a range $\mathcal{B}_n = [c_{min} b_n, c_{max} b_n]$ with
$c_{max}>c_{min}>0$ some constants and $b_n \downarrow 0$ at a
suitable rate. Alternatively one can use different bandwidths cells
for the $\check{\Sigma}_t$'s (see Patilea and Raïssi (2012) for more
details). The results in this paper are given uniformly with respect
to $b\in\mathcal{B}_n$. This justifies the approach which consists
in selecting the bandwidth on a grid defined in a range using for
example the cross validation criterion. Note also that the
$\check{\Sigma}_t$'s are positive definite. Of course our results do
not rely on a particular bandwidth choice procedure and are valid
provided estimators of the $\Sigma_t$'s with similar asymptotic
properties of the $\check{\Sigma}_t$'s are available. The non
parametric estimator of the covariance matrices employed in this paper is
similar to the variance estimators used
in Xu and Phillips (2008) among others. 
Considering the $\check{\Sigma}_t$'s, we are in
position to introduce the ALS estimators

\begin{equation*}
\hat{\theta}_{ALS}=\check{\Sigma}_{\tilde{X}}
^{-1}\mbox{vec}\:\left(\check{\Sigma}_{X}\right),
\end{equation*}
with

$$\check{\Sigma}_{\tilde{X}}=n^{-1}\sum_{t=1}^n\tilde{X}_{t-1}^p(\tilde{X}_{t-1}^{p})'
\otimes\check{\Sigma}_t^{-1}\quad\mbox{and}\quad\check{\Sigma}_{X}=n^{-1}
\sum_{t=1}^n\check{\Sigma}_t^{-1}X_t(\tilde{X}_{t-1}^{p})'.$$\\


Now we have to state the asymptotic behavior of the estimators and
introduce some notations. Define  $$\Delta=\left(
            \begin{array}{cccc}
              A_1 & \dots & A_{p-1} & A_p \\
              I_d & 0 & \dots & 0 \\
                & \ddots & \ddots & \vdots \\
              0 &  & I_d & 0 \\
            \end{array}
          \right)$$
and $e_p(1)$ the vector of dimension $p$ such that the first
component is equal to one and zero elsewhere. Under {\bf A1} it is
shown in Patilea and Raïssi (2012) that

\begin{equation}\label{res2}
\sqrt{n}(\hat{\theta}_{OLS}-\theta_0)\Rightarrow
\mathcal{N}(0,\Lambda_3^{-1}\Lambda_2\Lambda_3^{-1}),
\end{equation}
with $$\Lambda_2=\int_0^1
\sum_{i=0}^{\infty}\left\{\Delta^i(e_p(1)e_p(1)'
\otimes\Sigma(r))\Delta^{i'}\right\}\otimes\Sigma(r)\:dr,$$
$$\Lambda_3=\int_0^1
\sum_{i=0}^{\infty}\left\{\Delta^i(e_p(1)e_p(1)'
\otimes\Sigma(r))\Delta^{i'}\right\}\otimes I_d\:dr,$$ and
\begin{equation}\label{res1}
\sqrt{n}(\hat{\theta}_{GLS}-\theta_0)\Rightarrow
\mathcal{N}(0,\Lambda_1^{-1}),
\end{equation}
where  $$\Lambda_1=\int_0^1
\sum_{i=0}^{\infty}\left\{\Delta^i(e_p(1)e_p(1)'
\otimes\Sigma(r))\Delta^{i'}\right\}\otimes\Sigma(r)^{-1}\:dr.$$ In
addition we may use the following consistent estimators for the
covariance matrices: {\small
\begin{equation}\label{refestim}
\check{\Sigma}_{\tilde{X}}=\Lambda_1+o_p(1),\:\hat{\Sigma}_{\tilde{X}}=\Lambda_3+o_p(1)
\quad\mbox{and}\quad\hat{\Lambda}_2:=
n^{-1}\sum_{t=1}^n\tilde{X}_{t-1}^p\tilde{X}_{t-1}^{p'}\otimes
\hat{u}_t\hat{u}_t'=\Lambda_2+o_p(1).
\end{equation}}

We make the following assumptions to state the asymptotic equivalence between the ALS and GLS estimators.\\

\textbf{Assumption A1':} Suppose that all the conditions in
Assumption \textbf{A1} hold true.  In addition $\inf_{r\in(0,1]} \lambda_{min}(\Sigma(r)) >0$ where for any symmetric matrix $M$ the real value $ \lambda_{min}(M)$ denotes its smallest eigenvalue.


\vspace{0.3 cm}

\textbf{Assumption A2:} \, (i) The kernel $K(\cdot)$ is a bounded
density function defined on the real line such that $K(\cdot)$ is
nondecreasing on $(-\infty, 0]$ and decreasing on $[0,\infty)$ and
$\int_\mathbb{R} v^2K(v)dv < \infty$. The function $K(\cdot)$ is
differentiable except a finite number of points and the derivative
$K^\prime(\cdot)$  is an integrable function. Moreover, the Fourier
Transform $\mathcal{F}[K](\cdot)$ of $K(\cdot)$ satisfies
$\int_{\mathbb{R}}  \left| s \mathcal{F}[K](s) \right|ds <\infty$.

(ii) The bandwidth $b$ is taken in
the range $\mathcal{B}_n = [c_{min} b_n, c_{max} b_n]$ with $0<
c_{min}< c_{max}< \infty$ and $b_n + 1/Tb_n^{2+\gamma} \rightarrow
0$ as $n\rightarrow \infty$, for some $\gamma >0$.

(iii) The sequence $\nu_n$ is such that $n\nu_n^2 \rightarrow 0.$\\

Under these additional assumptions Patilea and Raïssi (2012) also
showed that

\begin{equation}\label{res3}
\sqrt{n}(\hat{\theta}_{ALS}-\hat{\theta}_{GLS})=o_p(1),
\end{equation}
and
\begin{equation}\label{res4}
\check{\Sigma}_{[Tr]} \stackrel{P}{\longrightarrow}
\Sigma(r-)\int_{-\infty}^0 K(z) dz + \Sigma(r+)\int_0^\infty K(z)
dz,
\end{equation}
where $\Sigma(r-):=\lim_{\tilde{r}\uparrow r}\Sigma(\tilde{r})$ 
and $\Sigma(r+):=\lim_{\tilde{r}\downarrow r}\Sigma(\tilde{r})$.
As a consequence $\hat{\theta}_{ALS}$ and $\hat{\theta}_{GLS}$ have
the same asymptotic behavior and we can also write
$\check{\Sigma}_{t}=\Sigma_t+o_p(1)$,
unless at the break dates where we have
$\check{\Sigma}_{t}=\Sigma_t+O_p(1)$. Using these asymptotic results
we underline the unreliability of the standard $AIC$ and develop a
criterion which is adapted to the case of non stationary but stable
processes. Corrected confidence bounds for the PAM and PCM in our
non standard framework are also proposed.

%
%

\section{Derivation of the adaptive AIC}
\label{S3}

In the standard case (the variance of the innovations is constant
with true variance $\Sigma_u$) the Kullback-Leibler discrepancy
between the true model and the approximating model with parameter
vector $\hat{\theta}_{OLS}$ is given by

\begin{equation}\label{dux}
d_n(\hat{\theta}_{OLS},\theta_0)=E_{\theta_0,\Sigma_u}
\left\{-2\mathcal{L}_{OLS}(\theta,\Sigma_u)\right\}\mid_{\theta=\hat{\theta}_{OLS}},
\end{equation}
see Brockwell and Davis (1991, p 302). Akaike (1973) proposed the
following approximately unbiased estimator of (\ref{dux}) to compare
the discrepancies between competing VAR($p$) models

$$AIC(p)=-2\mathcal{L}_{OLS}(\hat{\theta}_{OLS},\hat{\Sigma}_u)+\frac{2pd^2}{n},$$
where the term $2pd^2$ penalizes the more complicated models fitted
to the data (see Lütkepohl (2005), p 147). The terms corresponding
to the nuisance parameters are neglected in the previous expressions
since they do not interfere in the model selection when the $AIC$ is
used. The identified model corresponds to the model which minimizes
the $AIC$.
%
However in our non standard framework it is clear that the
$\mathcal{L}_{OLS}$ cannot take into account the non constant
variance in the observations. In addition if we assume that the
variance of the innovations is constant $\Sigma_t=\Sigma_u$, we
obtain

\begin{equation}\label{resstd}
\sqrt{n}(\hat{\theta}_{OLS}-\theta_0)\Rightarrow
\mathcal{N}(0,\Lambda_4^{-1}),
\end{equation}
with
$\Lambda_4=E(\tilde{X}_{t-1}\tilde{X}_{t-1}')\otimes\Sigma_u^{-1},$
so that the following result is used for the derivation of the
standard $AIC$

$$E_{\theta_0,\Sigma_u}\left\{n(\hat{\theta}_{OLS}-\theta_0)'\hat{\Lambda}_4(\hat{\theta}_{OLS}-\theta_0)\right\}\approx pd^2,$$
for large $n$, where $\hat{\Lambda}_4=n^{-1}\sum_{t=1}^n\tilde{X}_{t-1}^p(\tilde{X}_{t-1}^{p})'\otimes\hat{\Sigma}_u^{-1}$ is a consistent estimator of $\Lambda_4$. In view of (\ref{res2}) this property is obviously not verified in our case. Indeed Patilea and Raïssi (2012) pointed out that $\Lambda_4$ and $\Lambda_3^{-1}\Lambda_2\Lambda_3^{-1}$ can be quite different. Therefore the standard $AIC$ have no theoretical basis in our non standard framework and we can expect that the use of the standard $AIC$ can be misleading in such a situation.\\

To remedy to this problem we shall use the more appropriate
expression (\ref{likeals}) in our framework for the Kullback-Leibler
discrepancy between the fitted model and the true model

\begin{equation}\label{duxgls}
\Delta_n(\hat{\theta}_{GLS},\theta_0)=E_{\theta_0}\left\{-2\mathcal{L}_{GLS}(\theta)\right\}\mid_{\theta=\hat{\theta}_{GLS}}.
\end{equation}
Using a second order Taylor expansion of $\mathcal{L}_{GLS}$ about
$\hat{\theta}_{GLS}$ and since
$\frac{\mathcal{L}_{GLS}(\hat{\theta}_{GLS})}{\partial \theta}=0$,
we obtain

\begin{eqnarray}
E_{\theta_0}\left\{-2\mathcal{L}_{GLS}(\theta_0)\right\}&=&
\frac{1}{n}E_{\theta_0}\left\{n(\hat{\theta}_{GLS}-\theta_0)'\hat{\Sigma}_{\underline{\tilde{X}}}(\hat{\theta}_{GLS}-\theta_0)\right\}
\nonumber\\&&+
E_{\theta_0}\left\{-2\mathcal{L}_{GLS}(\hat{\theta}_{GLS})\right\}+o(1).\label{eq1}
\end{eqnarray}
Using again the second order Taylor expansion and taking the
expectation we also write
\begin{eqnarray}
E_{\theta_0}\left[E_{\theta_0}\left\{-2\mathcal{L}_{GLS}(\theta)\right\}\mid_{\theta=\hat{\theta}_{GLS}}\right]
&=&\frac{1}{n}E_{\theta_0}\left[n(\hat{\theta}_{GLS}-\theta_0)'E_{\theta_0}\left\{
\hat{\Sigma}_{\underline{\tilde{X}}}\right\}(\hat{\theta}_{GLS}-\theta_0)\right]
\nonumber\\&&+E_{\theta_0}\left\{-2\mathcal{L}_{GLS}(\theta_0)\right\}\label{eq2}
+o(1).
\end{eqnarray}
From (\ref{res1}) we have for large $n$

$$E_{\theta_0}\left[n(\hat{\theta}_{GLS}-\theta_0)'\hat{\Sigma}_{\underline{\tilde{X}}}(\hat{\theta}_{GLS}-\theta_0)\right]\approx pd^2$$
and

$$E_{\theta_0}\left[n(\hat{\theta}_{GLS}-\theta_0)'E_{\theta_0}\left\{
\hat{\Sigma}_{\underline{\tilde{X}}}\right\}(\hat{\theta}_{GLS}-\theta_0)\right]\approx
pd^2.$$ Noting that

\begin{eqnarray*}
&&E_{\theta_0}\left[E_{\theta_0}\left\{-2\mathcal{L}_{GLS}(\theta)\right\}\mid_{\theta=\hat{\theta}_{GLS}}\right]=
E_{\theta_0}\left\{-2\mathcal{L}_{GLS}(\hat{\theta}_{GLS})\right\}
\\&&+E_{\theta_0}\left[E_{\theta_0}\left\{-2\mathcal{L}_{GLS}(\theta)\right\}\mid_{\theta=\hat{\theta}_{GLS}}\right]
-E_{\theta_0}\left\{-2\mathcal{L}_{GLS}(\theta_0)\right\}
\\&&+E_{\theta_0}\left\{-2\mathcal{L}_{GLS}(\theta_0)\right\}-
E_{\theta_0}\left\{-2\mathcal{L}_{GLS}(\hat{\theta}_{GLS})\right\}
\end{eqnarray*}
and using (\ref{eq1}) and (\ref{eq2}), we see that the following
criterion based on the GLS estimator
$$AIC_{GLS}=-2\mathcal{L}_{GLS}(\hat{\theta}_{GLS})+\frac{2pd^2}{n}$$
is an approximately unbiased estimator of
$\Delta_n(\hat{\theta}_{GLS},\theta_0)$.

Nevertheless the $AIC_{GLS}$ is infeasible since it depends on the
unknown variance of the errors. Thereby we will use the adaptive
estimation of the variance structure to propose a feasible selection
criterion. Recall that

\begin{equation*}
-2\mathcal{L}_{GLS}(\hat{\theta}_{GLS})=
\frac{1}{n}\sum_{t=1}^n\ln\left\{\det(\Sigma_t)\right\}+u_t(\hat{\theta}_{GLS})'\Sigma_t^{-1}u_t(\hat{\theta}_{GLS}),
\end{equation*}
and define

\begin{equation*}
-2\mathcal{L}_{ALS}(\hat{\theta}_{ALS})=
\frac{1}{n}\sum_{t=1}^n\ln\left\{\det(\check{\Sigma}_t)\right\}+u_t(\hat{\theta}_{ALS})'\check{\Sigma}_t^{-1}u_t(\hat{\theta}_{ALS}).
\end{equation*}
In view of (\ref{res3}) and (\ref{res4}) we have

$$-2\mathcal{L}_{GLS}(\hat{\theta}_{GLS})=-2\mathcal{L}_{ALS}(\hat{\theta}_{ALS})+o_p(1),$$
since we allowed for a finite number of variance breaks for the
innovations. Therefore we can introduce the adaptive criterion

$$AIC_{ALS}=-2\mathcal{L}_{ALS}(\hat{\theta}_{ALS})+\frac{2pd^2}{n},$$
which gives an approximately unbiased estimation of (\ref{duxgls})
for large $n$.

Finally note that if we suppose that $(X_t)$ is cointegrated, Kim
and Park (2010) showed that the long run relationships estimated by
reduced rank are $n$-consistent.
Therefore our approach for building information criteria can be
straightforwardly extended to the cointegrated case since it is
clear that the estimated long run relationships can be replaced by
the true relationships in the preceding computations.

\section{Identifying the lag length using partial autoregressive and partial correlation matrices}
\label{Spartial}

In this part we assume $p>p_0$, so that the cut-off property of the
presented tools can be observed. Following the approach described in
Reinsel (1993) chapter 3, one can use the estimators of the
autoregressive parameters to identify the lag length of (\ref{VAR}).
Consider the regression of $X_t$ on its past values
\begin{eqnarray}\label{VARk}
{X}_t={A}_{01}{X}_{t-1}+\dots+{A}_{0p}{X}_{t-p}+u_t.
\end{eqnarray}
We can remark that the partial autoregressive matrices
$A_{0p_0+1},\dots,A_{0p}$ are equal to zero. The PAM are estimated
using OLS or ALS estimation. Confidence bounds for the PAM can be
proposed
as follow. 
%
%
Let us introduce the $d^2(p-p_0)\times d^2p$ dimensional matrix
$R=(0,I_{d^2(p-p_0)})$, so that from (\ref{res2}), (\ref{res1}) and
(\ref{res3}) we write

\begin{equation}\label{respartial}
\sqrt{n}R\hat{\theta}_{OLS}\Rightarrow
\mathcal{N}(0,R\Lambda_{3}^{-1}\Lambda_{2}\Lambda_{3}^{-1}R')\quad\mbox{and}\quad\sqrt{n}R\hat{\theta}_{ALS}\Rightarrow
\mathcal{N}(0,R\Lambda_{1}^{-1}R'),
\end{equation}
where $R\hat{\theta}_{OLS}$ and $R\hat{\theta}_{ALS}$ correspond to
the OLS and ALS estimators of the null matrices
$A_{0p_0+1},\dots,A_{0p}$. Denote by $\upsilon_{i}^{OLS}$ (resp.
$\upsilon_{i}^{ALS}$) the asymptotic standard deviation of the $i$th
component of $\hat{\theta}_{OLS}$ (resp. $\hat{\theta}_{ALS}$) for
$i\in\{d^2p_0+1,\dots,d^2p\}$ with obvious notations. From
(\ref{respartial}) the $i$-th component of $\hat{\theta}_{OLS}$
(resp. $\hat{\theta}_{ALS}$) are usually compared with the 95\%
approximate asymptotic confidence bounds
$\pm1.96\hat{\upsilon}_{i}^{OLS}$ (resp.
$\pm1.96\hat{\upsilon}_{i}^{ALS}$) as suggested in Tiao and Box
(1981),
and where the $\hat{\upsilon}_{i}^{OLS}$'s and $\hat{\upsilon}_{i}^{ALS}$'s
can be obtained using the consistent estimators in (\ref{refestim}).
Therefore the identified lag length for model (\ref{VAR}) correspond
to the higher order of the matrix $A_{0i}$
which have an estimator of a component which is clearly beyond its confidence bounds about zero.\\

The identification of the lag length of standard VAR processes is
usually performed using also the partial cross-correlation matrices
which are the extension of the partial correlations of the
univariate case. Consider the regressions
$$X_{t-p}=\phi_1X_{t-p+1}+\dots+\phi_{p-1}X_{t-1}+w_t,$$

\begin{equation*}
{X}_t={A}_{01}{X}_{t-1}+\dots+{A}_{0p-1}{X}_{t-p+1}+u_t,
\end{equation*}
with $p>1$. In our framework it is clear that the error process
$(w_t)$ is unconditionally heteroscedastic, and then
we define $\underline{\Sigma}_w=\lim_{n\to\infty}n^{-1}\sum_{t=1}^nE(w_tw_t')$ 
which converge under {\bf A1}, and the consistent estimator of
$\underline{\Sigma}_w$:

\begin{eqnarray*}
\hat{\underline{\Sigma}}_w&=&n^{-1}\sum_{t=p}^nX_{t-p}X_{t-p}'
\\&-&\left\{n^{-1}\sum_{t=p}^nX_{t-p}(\tilde{X}_{t-p}^{p-1})'
\right\}\left\{n^{-1}\sum_{t=p}^n\tilde{X}_{t-p}^{p-1}(\tilde{X}_{t-p}^{p-1})'\right\}^{-1}
\left\{n^{-1}\sum_{t=p}^n\tilde{X}_{t-p}^{p-1}X_{t-p}'\right\}
\end{eqnarray*}
with obvious notations. The consistency of this estimator can be
proved from standard computations and using lemmas 7.1-7.4 of
Patilea and Raïssi (2012). We also define the 'long-run' innovations
variance
$\underline{\Sigma}_u=\lim_{n\to\infty}n^{-1}\sum_{t=1}^nE(u_tu_t')$
where the $E(u_tu_t')$'s are non constant and the consistent
estimator $\hat{\underline{\Sigma}}_u=n^{-1}\sum_{t=1}^n
\hat{u}_t\hat{u}_t'$ of $\underline{\Sigma}_u$ where we recall that
the $\hat{u}_t$'s are the OLS residuals.

Several definitions for the partial cross-correlations are available
in the literature. In the sequel we concentrate on the definition
given in Ansley and Newbold (1979) which is used in the VARMAX
procedure of the software SAS. We propose to extend the partial
cross-correlation matrices in our framework as follow

\begin{equation}\label{eqzero}
P(p)=\left(\underline{\Sigma}_u^{-\frac{1}{2}}\otimes\underline{\Sigma}_w^{-\frac{1}{2}}\right)
\mbox{vec}\left\{n^{-1}\sum_{t=1}^nE(w_tu_t')\right\}
=\left(\underline{\Sigma}_u^{-\frac{1}{2}}\otimes\underline{\Sigma}_w^{\frac{1}{2}}\right)\mbox{vec}(A_p)
\end{equation}
and it is clear that for $p>p_0$ we have $P(p)=0$. The expression
(\ref{eqzero}) may be viewed as the 'long-run' relation between the
$X_t$'s and the $X_{t-p}$'s corrected for the intermediate values
for each date $t$. Consider the OLS and ALS consistent estimators

$$\hat{P}_{OLS}(p)=\left(\hat{\underline{\Sigma}}_u^{-\frac{1}{2}}\otimes
\hat{\underline{\Sigma}}_w^{\frac{1}{2}}\right)\mbox{vec}(\tilde{R}\hat{\theta}_{OLS})$$

$$\hat{P}_{ALS}(p)=\left(\hat{\underline{\Sigma}}_u^{-\frac{1}{2}}\otimes
\hat{\underline{\Sigma}}_w^{\frac{1}{2}}\right)\mbox{vec}(\tilde{R}\hat{\theta}_{ALS}),$$
where $\tilde{R}=(0,I_{d^2})$ is of dimension $d^2\times d^2p$, so
that $\tilde{R}\hat{\theta}_{OLS}$ and $\tilde{R}\hat{\theta}_{ALS}$
correspond to the ALS and OLS estimators of $A_p$ in (\ref{VARk}).
Using again (\ref{res2}), (\ref{res1}), (\ref{res3}) and from the
consistency of $\hat{\underline{\Sigma}}_u$ and
$\hat{\underline{\Sigma}}_w$ we obtain

\begin{equation}\label{eqone}
n^{\frac{1}{2}}\hat{P}_{OLS}(p)\Rightarrow\mathcal{N}\left(0,\left(\underline{\Sigma}_u^{-\frac{1}{2}}\otimes
\underline{\Sigma}_w^{\frac{1}{2}}\right)\left(\tilde{R}\Lambda_{3}^{-1}\Lambda_{2}\Lambda_{3}^{-1}\tilde{R}'\right)
\left(\underline{\Sigma}_u^{-\frac{1}{2}}\otimes
\underline{\Sigma}_w^{\frac{1}{2}}\right)\right)
\end{equation}

\begin{equation}\label{eqtwo}
n^{\frac{1}{2}}\hat{P}_{ALS}(p)\Rightarrow\mathcal{N}\left(0,\left(\underline{\Sigma}_u^{-\frac{1}{2}}\otimes
\underline{\Sigma}_w^{\frac{1}{2}}\right)\left(\tilde{R}\Lambda_{1}^{-1}\tilde{R}'\right)\left(\underline{\Sigma}_u^{-\frac{1}{2}}\otimes
\underline{\Sigma}_w^{\frac{1}{2}}\right)\right).
\end{equation}
Hence approximate confidence bounds can be built using (\ref{eqone})
and (\ref{eqtwo}). Similarly to the partial autoregressive matrices
the highest order $p$ for which a cut-off is observed for an element
of $\hat{P}_{OLS}(p)$ ( resp. $\hat{P}_{ALS}(p)$) correspond to the
identified lag length for the VAR model. Note that for $p=1$
($p_0=0$, so that the observed process is uncorrelated and
$w_t=X_{t-1}$, $u_t=X_t$), we have
$\lim_{n\to\infty}n^{-1}\sum_{t=1}^nE(X_tX_t')=\underline{\Sigma}_w=\underline{\Sigma}_u$.
In this case similar results to (\ref{eqone}) and (\ref{eqtwo}) can be used.\\
%
%

Let us end this section with some remarks on the OLS and ALS
estimation approaches of the PAM and PCM. If we assume that the
variance of the error process is constant, the result (\ref{resstd})
is used to identify the autoregressive order using the partial
autoregressive and correlation matrices obtained from the OLS
estimation. However as pointed out in the previous section this
standard result can be misleading in our framework. The simulations carried out in the next section show that the standard bounds are not reliable in our framework. From the real
example below it appears that the OLS PAM and PCM with the standard
confidence bounds seem to select a too large lag length. Note that
since the tools presented in this section are based on the results
(\ref{res2}), (\ref{res1}) and on the adaptive estimation of the
autoregressive parameters, they are able to take into account
changes in the variance.

In the univariate case the partial autocorrelation function is used
for identifying the autoregressive order. In such a case the
asymptotic behavior of $\hat{\theta}_{ALS}$ does not depend on the
variance structure (see equation (\ref{phixu}) below). Hence the
ALS estimators of the partial autocorrelations do not depend on the
variance function on the contrary to its OLS counterparts. In the
general VAR case Patilea and Raïssi (2012) also showed that that
$\Lambda_3^{-1}\Lambda_2\Lambda_3^{-1} - \Lambda_1^{-1} $ is
positive semi-definite (the same result is available in the
univariate case). Therefore the tools based on the ALS estimator are
more accurate than the tools based on the OLS estimator for
identifying the autoregressive order.

We illustrate the above remarks by considering the following simple
case where we assume that $\Sigma(r)=\sigma(r)^2I_d$ with
$\sigma(\cdot)^2$ a real-valued function. The univariate stable
autoregressive processes are a particular case of this specification
of the variance. In this case we obtain
\begin{equation}\label{phixu} \Lambda_1=\Lambda_4=
\sum_{i=0}^{\infty}\left\{\Delta^i(e_p(1)e_p(1)'\otimes
I_d)\Delta^{i'}\right\}\otimes I_d
\end{equation}
and
\begin{equation*}\Lambda_2=\int_0^1\sigma(r)^4dr
\Lambda_1,\qquad\Lambda_3=\int_0^1\sigma(r)^2dr \Lambda_1,
\end{equation*}
so that we have

\begin{equation}\label{lastequ}
\Lambda_3^{-1}\Lambda_2\Lambda_3^{-1}=
\frac{\int_0^1\sigma(r)^4dr}{\left(\int_0^1\sigma(r)^2dr\right)^2}\Lambda_1^{-1}
\end{equation}
with $\int_0^1\sigma(r)^4dr\geq\left(\int_0^1\sigma(r)^2dr\right)^2$.
Hence from (\ref{lastequ}) it is clear
that the adaptive PAM and PCM are more reliable than the PAM and PCM
obtained using the OLS approach. In addition from (\ref{phixu}) the
asymptotic behavior of the adaptive partial autoregressive and
partial correlation matrices does not depend on the variance
function. On the other hand we also see from (\ref{lastequ}) that
the matrices $\Lambda_4$ and $\Lambda_3^{-1}\Lambda_2\Lambda_3^{-1}$
can be quite different. 



\section{Empirical results}
\label{S4}

For our empirical study the $AIC_{ALS}$ is computed using an
adaptive estimation of the variance as described in Section
\ref{S2}. The ALS estimators of the PAM and PCM are obtained
similarly. In particular the bandwidth is selected using the
cross-validation method. The OLS partial autoregressive and partial
correlation matrices used with the standard confidence bounds are
denoted by
$PAM_S$ and $PCM_S$. Similarly we also introduce the $PAM_{OLS}$, $PAM_{ALS}$ 
 and
the $PCM_{OLS}$, $PCM_{ALS}$
with obvious notations. In the simulation study part the infeasible
and $GLS$ tools are used only for comparison with the feasible  
$ALS$ tools.

It is important to note that when the $PAM$ and $PCM$ are used, the
practitioners base their decision on the visual inspection of these
tools.
Results concerning
automatically selected lag lengths over the iterations using several
$PAM$ and $PCM$ do not really reflect their ability to identify the
lag length in practice. For instance it is well known that there are
cases where some $PAM$ or $PCM$ are beyond the confidence bounds but
not taken into account for the lag length identification. Therefore
we provide instead some simulation results which assess the ability
of the studied methods to provide reliable confidence bounds for the
choice of the lag length.
The use of the modified $PAM$ and $PCM$ is also illustrated in the
real data study below.

For a given tool we assume in our experiments that when the selected
autoregressive order is such that $p>5$, the model identification is
suspected to be not reliable. For instance the more complicated
models may appear not enough penalized by the information criterion,
or the number of estimated parameters becomes too large when
compared to the number of observations. In such situations the
practitioner is likely to stop the procedure.

\subsection{Monte Carlo experiments}

In this part $N=1000$ independent trajectories of bivariate VAR(2)
($p_0=2$) processes of length $n=50$, $n=100$ and $n=200$ are
simulated with autoregressive parameters given by

\begin{equation}\label{rasrmel}
A_{01}=\left(
           \begin{array}{cc}
             -0.4 & 0.1 \\
             0 & -0.7 \\
           \end{array}
         \right)
\quad\mbox{and}\quad A_{02}=\left(
           \begin{array}{cc}
             -0.6 & 0 \\
             0 & -0.3 \\
           \end{array}
         \right).
\end{equation}
Recall that the process $(\epsilon_t)$ is assumed iid standard
Gaussian.
Two kinds of non constant volatilities are used. 
When the variance smoothly change in time we consider the following
specification

\begin{equation}\label{smooth}
\Sigma(r)=\left(
        \begin{array}{cc}
          (1+\gamma_1 r)(1+\rho^2) & \rho(1+\gamma_1 r)^{\frac{1}{2}}(1+\gamma_2 r)^{\frac{1}{2}} \\
          \rho(1+\gamma_1 r)^{\frac{1}{2}}(1+\gamma_2 r)^{\frac{1}{2}} & (1+\gamma_2 r) \\
        \end{array}
      \right).
\end{equation}
In case of abrupt change the following variance specification is
used

\begin{equation}\label{break}
\Sigma(r)=\left(
        \begin{array}{cc}
          (1+f_1(r))(1+\rho^2) & \rho(1+f_1(r))^{\frac{1}{2}}\rho(1+f_2(r))^{\frac{1}{2}} \\
          \rho(1+f_2(r))^{\frac{1}{2}}(1+f_1(r))^{\frac{1}{2}} & (1+f_2(r))(1+\rho^2) \\
        \end{array}
      \right),
\end{equation}
with $f_i(r)=(\gamma_i-1)\mathbf{1}_{(r\geq1/2)}(r)$. In this case
we have a common variance break at the date $t=n/2$. In all the
experiments we take $\gamma_1=20$, $\gamma_2=\gamma_1/3$ and
$\rho=0.2$. Note that the autoregressive parameters as well as the
variance structure are inspired by the real data study below. More precisely the
autoregressive parameters in (\ref{rasrmel}) are taken close to the
two first adaptive PAM obtained for the government securities
and foreign direct investment system. In addition the ratio between
the first and last adaptive estimation of the residual variance of
the studied real data are of the same order of the ratio for the
residual variance of the simulated series.
The results are given in Tables \ref{tab1}-\ref{tab3} and
\ref{tab1pampcm}-\ref{tab3pampcm} for the variance specification
(\ref{smooth}) and in Tables \ref{tab1b}-\ref{tab3b} and
\ref{tab1bpampcm}-\ref{tab3bpampcm} when specification (\ref{break})
is used. To facilitate the comparison of the studied identification
tools, the most frequently selected lag length for
the studied information criteria and the correct order ($p_0=2$) are in bold type.

The small sample properties of the different information criteria
for selecting the autoregressive order is first analyzed. According
to Tables \ref{tab1}-\ref{tab3b}
we can remark that the $AIC_{ALS}$ and $AIC_{GLS}$
are selecting most frequently the true autoregressive order. However we note that the modified $AIC$
have a slight tendency to select $p>p_0$. On the other hand we can see that the classical $AIC$ selects
too large lags lengths in our framework.
This is in accordance with the fact that $AIC$ is not consistent
(see e.g. Paulsen (1984) or Hurvich and Tsai (1989)).
We also note that the frequency of
selected true lag length $p=p_0=2$ increase with $n$ for the
$AIC_{GLS}$ and $AIC_{ALS}$.
The infeasible $AIC_{GLS}$ provide slightly better results than the
$AIC_{ALS}$. As expected it can be seen that the difference between
the $AIC_{GLS}$ and $AIC_{ALS}$ seems more marked when the processes
display an abrupt variance change. Indeed note that from
(\ref{res4}) the variance is not consistently estimated at the
break dates.
Nevertheless such bias is divided by $n$, and we note that the
behavior of the $AIC_{GLS}$ and $AIC_{ALS}$ become similar as the
samples increase in all the studied cases.
According to our simulation results it appears that the adaptive
$AIC$ is more able to select the appropriate autoregressive order
than the standard $AIC$ when the underlying process is indeed a VAR
process.

Now we turn to the analysis of the results for the PAM and PCM in
Tables \ref{tab1pampcm}-\ref{tab3bpampcm}. Note that we used the 5\%
(asymptotic) confidence bounds in our study. From our results it
emerges that the standard bounds do not provide reliable tools for
the identification of the lag length when the variance is non
constant. It can be seen that the frequencies of PAM and PCM with lag greater than $p_0$ beyond
the standard bounds do not converge to 5\%. On the other hand it is
found that the PAM and PCM based on the OLS and adaptive approaches
give satisfactory results. Indeed the frequencies of PAM and
PCM with lag greater than $p_0$ beyond the standard and adaptive bounds converge to 5\%. As
above we can remark that the results when the variance is smooth are
better than the case where the variance exihibits an abrupt break.
When the PAM and PCM are equal to zero it seems that the adaptive
and OLS method give similar results. In accordance with the
theoretical the more accurate adaptive method is more able than the
OLS method to detect the significant PAM and PCM with lag smaller or equal to $p_0$. We can draw the
conclusion that the standard bounds have to be avoided in our non
standard framework. The modified adaptive and OLS methods give
reliable approaches for identifying the lag length of a VAR model
with non constant variance. It emerges that the more elaborated
adaptive approach is preferable.



\subsection{Real data study}

\begin{figure}[h]\!\!\!\!\!\!\!\!\!\!
\vspace*{9.8cm} 


\protect \includegraphics{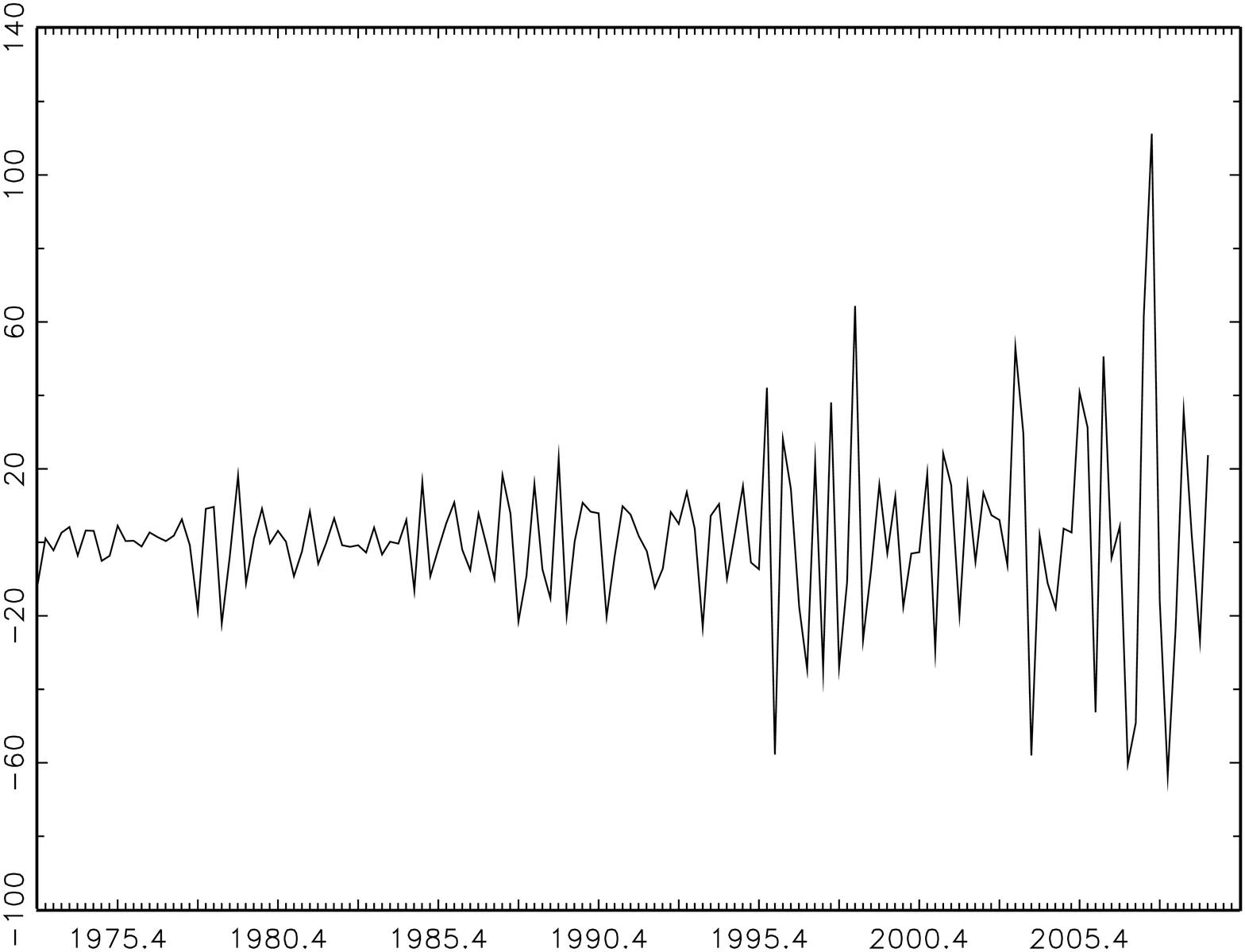} \protect \includegraphics{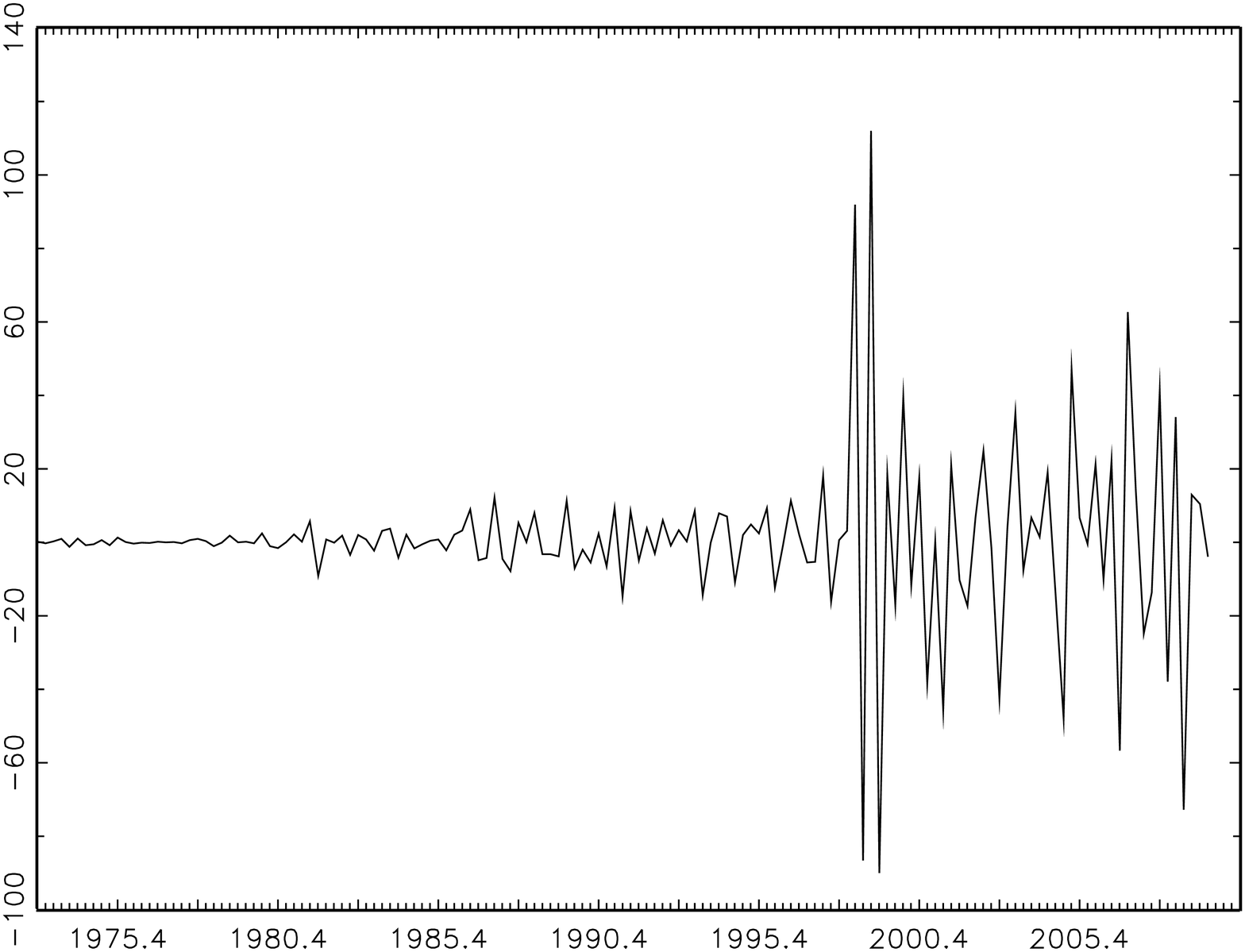} \caption{\label{datadiff} {\footnotesize The
differences of the government securities hold by foreigners (on the
left) and of the foreign direct investment (on the right) in
billions dollars $(n=147)$.}}
\end{figure}

In this part we try to identify the VAR order of a bivariate system
of variables taken from US international finance data. The first
differences of the quarterly US Government Securities (GS hereafter)
hold by foreigners and the Foreign Direct Investment (FDI hereafter)
in the US in billions dollars are studied from January 1, 1973 to
October 1, 2009. The length of the series is $n=147$. The studied
series are plotted in Figure \ref{datadiff} and can be downloaded
from website of the research division of the federal reserve bank of
Saint Louis:  www.research.stlouisfed.org.

\begin{figure}[h]\!\!\!\!\!\!\!\!\!\!
\vspace*{5.6cm} 


\protect \includegraphics{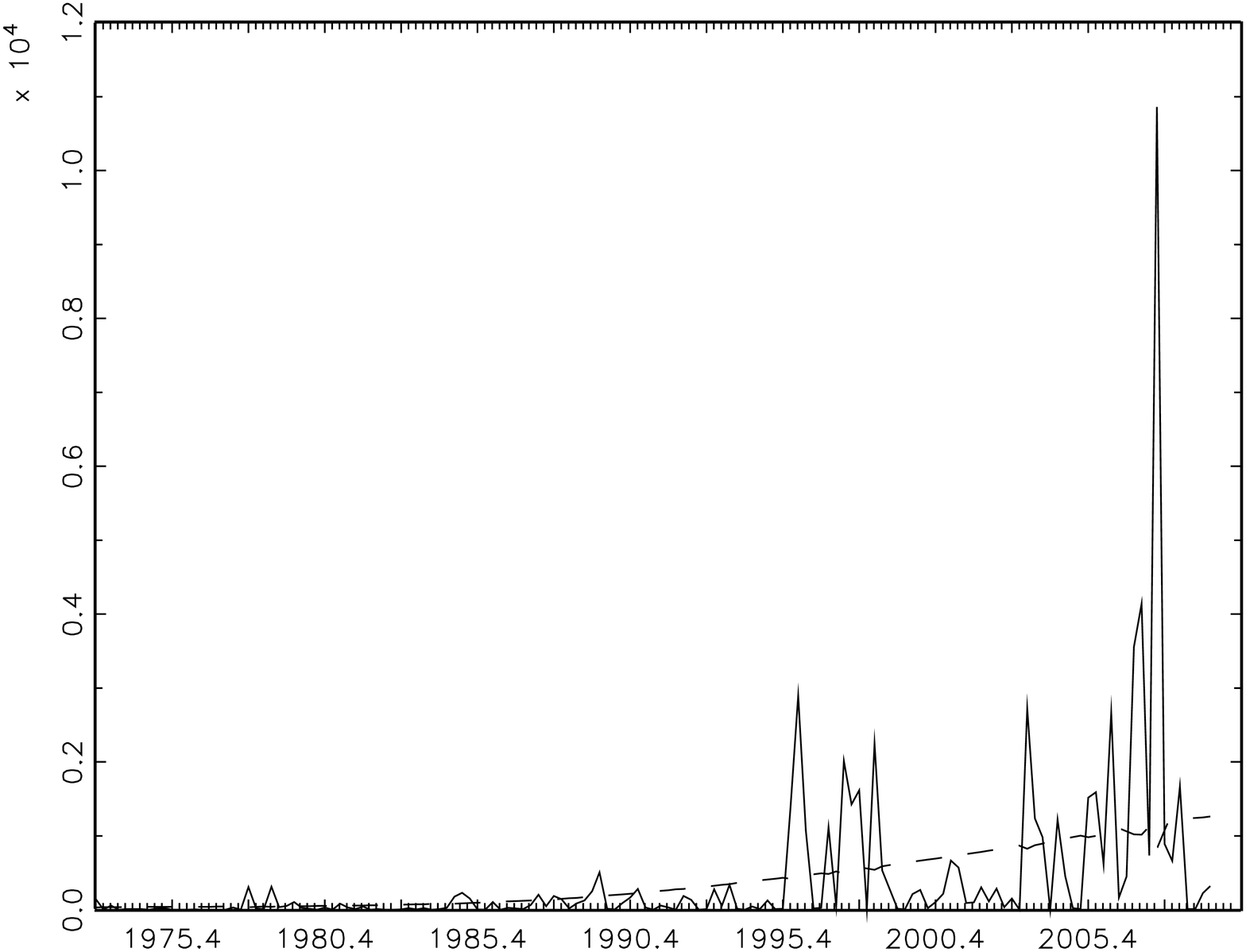} \protect \includegraphics{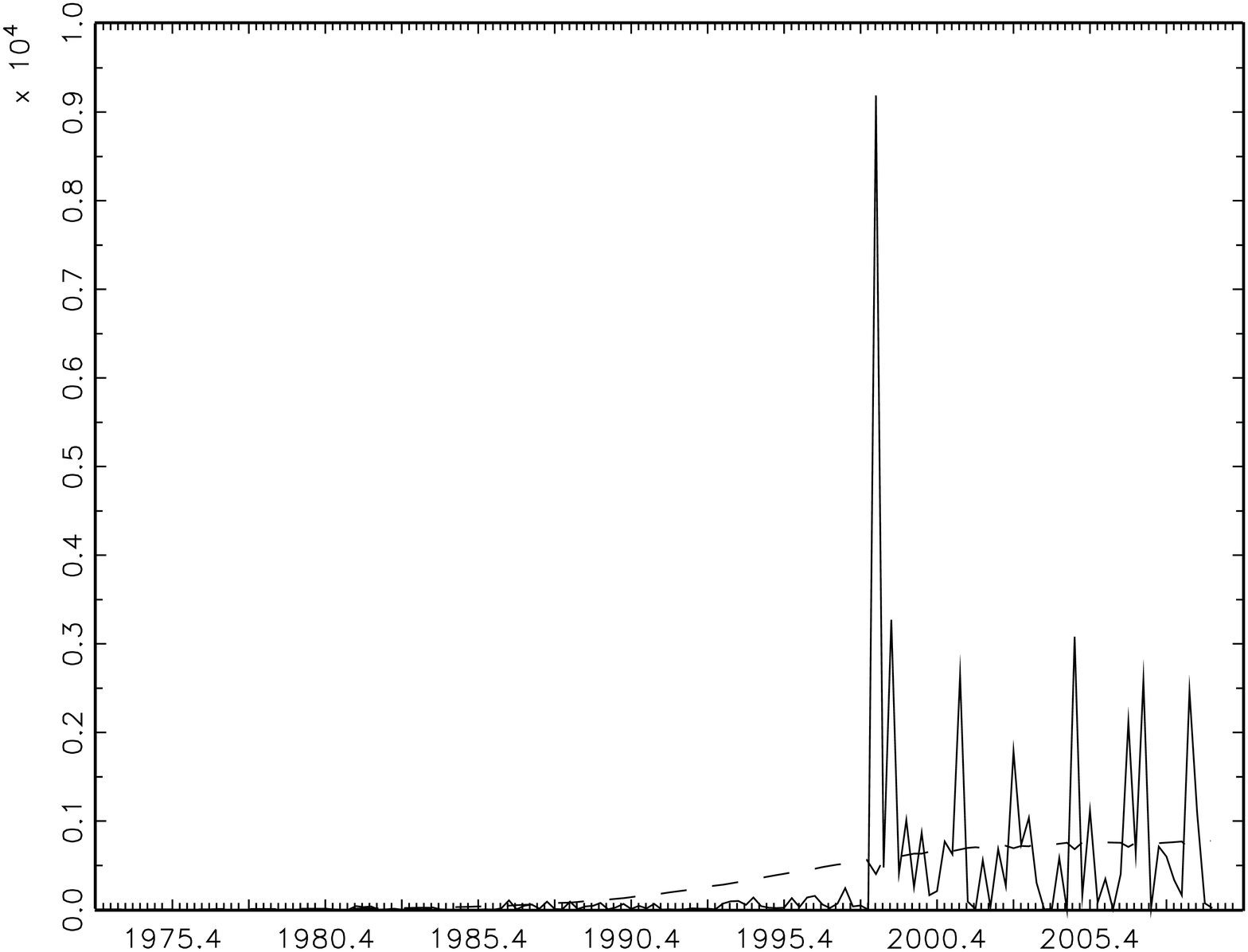}
\caption{\label{varestim} {\footnotesize The $\hat{u}_{1t}^2$'s
(full line) and the non parametric estimation of Var($u_{1t}$)
(dotted line) on the left and the same for the $\hat{u}_{2t}^2$'s
and Var($u_{2t}$) on the right.}}
\end{figure}

We first highlight some features of the studied series. The OLS
residuals and the variances of the errors estimated by kernel
smoothing are plotted in Figure \ref{varestim}. From Figure
\ref{datadiff} it appears that the data do not have a random walk
behavior,
while from Figure \ref{varestim} the estimated volatilities seem not constant. 
The residuals plotted in Figure \ref{varestim} show that the
variance of the first component of the residuals seems
constant from January 1973 to October 1995 and then we may suspect
an abrupt variance change. Similarly the variance of the second
component of the residuals seems constant from January 1973 to July
1998 and then we remark an abrupt variance change.
Therefore it clearly appears that the standard homoscedasticity
assumption turns out to be not realistic for the studied series.

\begin{figure}[h]\!\!\!\!\!\!\!\!\!\!
\vspace*{0.5cm} \hspace*{0.3cm}$\hat{P}_{OLS}^{11}(h)$
\hspace*{5.0cm}$\hat{P}_{OLS}^{12}(h)$

\vspace*{1.2cm} \hspace*{6.3cm}$h$ \hspace*{6.4cm}$h$
\vspace*{0.1cm}

\vspace*{0.9cm} \hspace*{0.3cm}$\hat{P}_{OLS}^{21}(h)$
\hspace*{5.1cm}$\hat{P}_{OLS}^{22}(h)$

\vspace*{1.2cm} \hspace*{6.3cm}$h$ \hspace*{6.4cm}$h$
\vspace*{0.8cm}

\protect \includegraphics{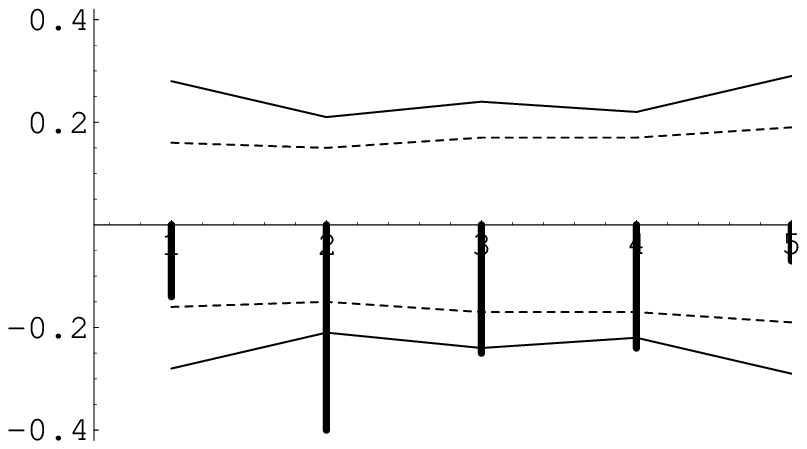} \protect \includegraphics{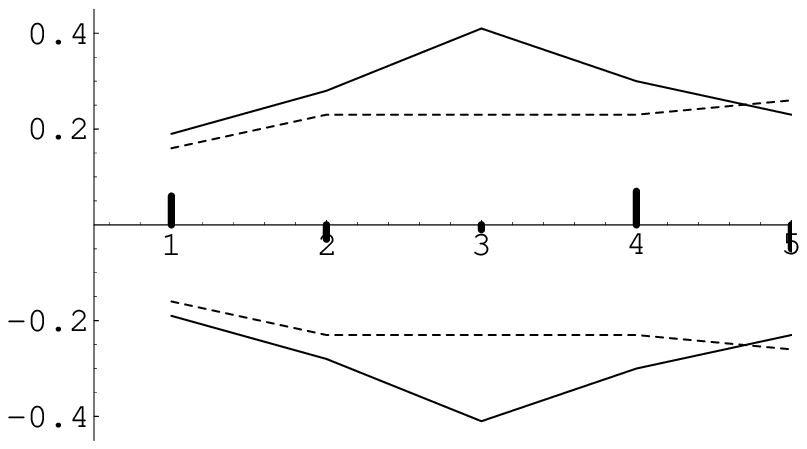} \protect
\includegraphics{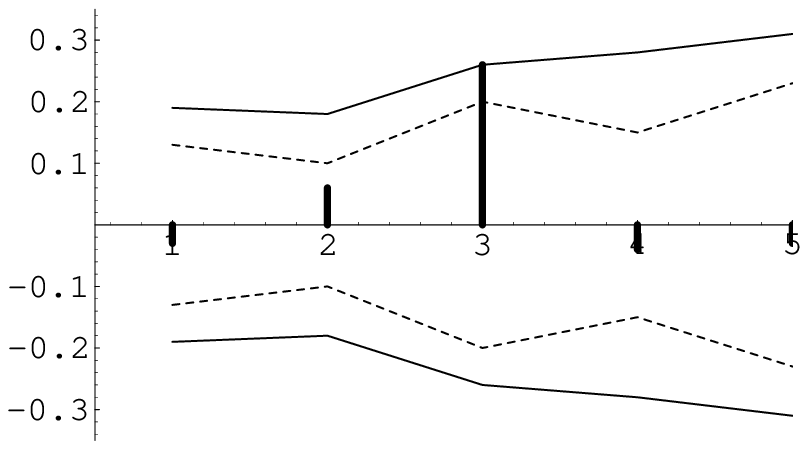} \protect \includegraphics{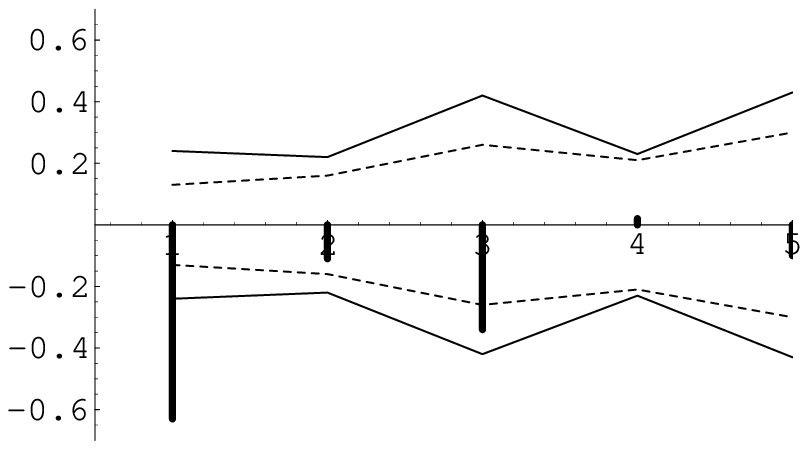} \caption{\label{autocorrols1}
{\footnotesize The OLS partial correlation matrices. The 95\% OLS
confidence bounds (in full lines) are obtained from (\ref{eqone})
while the 95\% standard confidence bounds (in dotted lines) are
obtained using (\ref{resstd}).}}
\end{figure}

\begin{figure}[h]\!\!\!\!\!\!\!\!\!\!
\vspace*{0.5cm} \hspace*{0.3cm}$\hat{P}_{ALS}^{11}(h)$
\hspace*{5.0cm}$\hat{P}_{ALS}^{12}(h)$

\vspace*{1.2cm} \hspace*{6.3cm}$h$ \hspace*{6.4cm}$h$
\vspace*{0.1cm}

\vspace*{0.9cm} \hspace*{0.3cm}$\hat{P}_{ALS}^{21}(h)$
\hspace*{5.1cm}$\hat{P}_{ALS}^{22}(h)$

\vspace*{1.2cm} \hspace*{6.3cm}$h$ \hspace*{6.4cm}$h$
\vspace*{0.8cm}

\protect \includegraphics{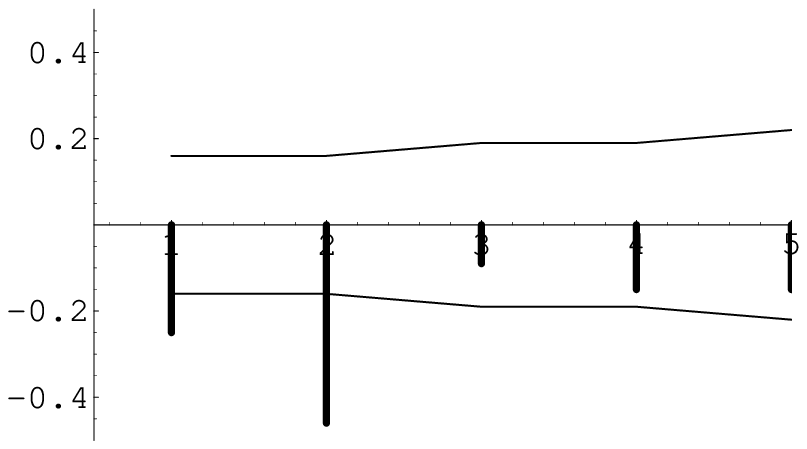} \protect \includegraphics{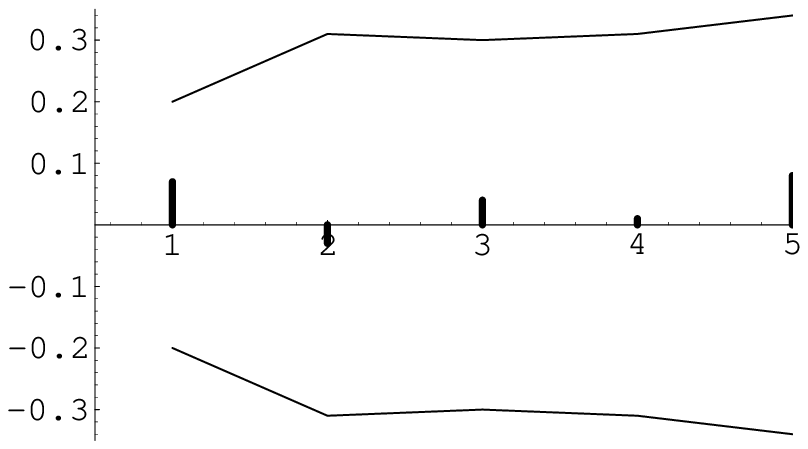} \protect
\includegraphics{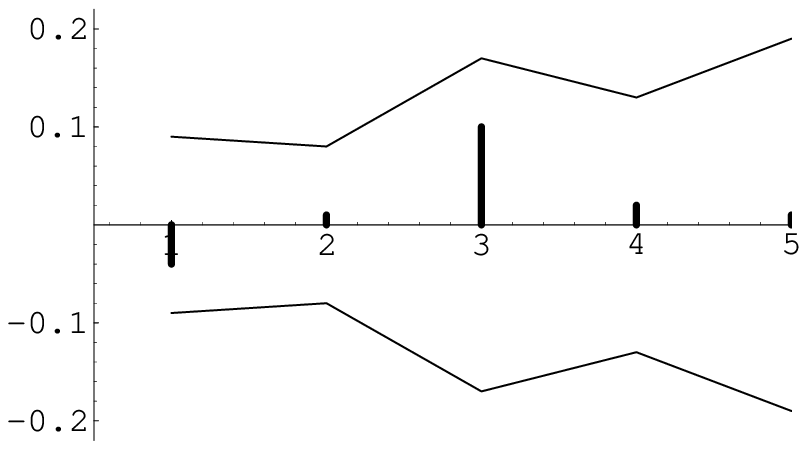} \protect \includegraphics{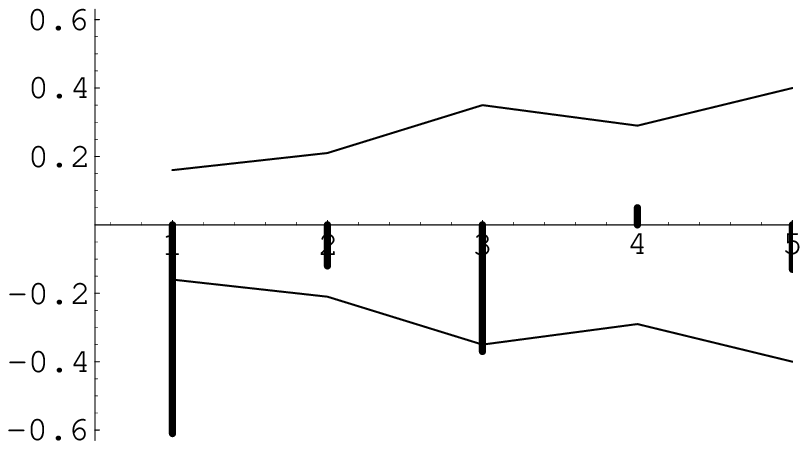} \caption{\label{autocorrols2}
{\footnotesize The ALS partial correlation matrices. The 95\%
confidence bounds are obtained from (\ref{eqtwo})}}
\end{figure}

We fitted VAR($p$) models with $p\in\{1,\dots,5\}$ to the data and
computed the $AIC$ and $AIC_{ALS}$ for each $p$. In our VAR system
the first component corresponds to the GS and the second corresponds
to the FDI. From Table \ref{realdata} the $AIC$ is decreasing as $p$
is increased so that the higher autoregressive order $p=5$ is
selected, while the minimum value for the $AIC_{ALS}$ is attained
for $p=2$. If it is assumed that the studied processes follow a VAR
model and since we noted that the variance of the studied processes
seems non constant, it is likely that the $AIC$ is not reliable and
selects a too large autoregressive order. In view of our above
results the model identification with the more parsimonious
$AIC_{ALS}$ seems more reliable.
We also considered the $PCM$ obtained from the standard, OLS and ALS
estimation methods. The $PCM$ are plotted in Figures
\ref{autocorrols1} and \ref{autocorrols2} and it appear that we can
identify $p=2$ using the modified tools while $p=3$ could be
identified using the standard $PCM$. We also see that the standard
and OLS confidence bounds can be quite different. The PAM are given
below with the 95\% confidence bounds into brackets. We base our lag
length choice on the $PAM$ which are clearly greater than its 95\%
confidence bounds (in bold type). The $PAM_S$ give for the studied
data:

$$\hat{A}_1^S=\left(
          \begin{array}{cc}
            {\bf -0.35}_{[\pm 0.16]} & 0.12_{[\pm 0.19]} \\
            0.06_{[\pm 0.14]} & {\bf -0.72}_{[\pm 0.16]} \\
          \end{array}
        \right)\quad\hat{A}_2^S=\left(
          \begin{array}{cc}
            {\bf -0.55}_{[\pm 0.17]} & 0.08_{[\pm 0.23]} \\
            0.08_{[\pm 0.14]} & {\bf -0.27}_{[\pm 0.20]} \\
          \end{array}
        \right)$$

$$\hat{A}_3^S=\left(
          \begin{array}{cc}
            {\bf -0.32}_{[\pm 0.18]} & 0.07_{[\pm 0.23]} \\
            0.15_{[\pm 0.16]} & -0.20_{[\pm 0.20]} \\
          \end{array}
        \right)\quad\hat{A}_4^S=\left(
          \begin{array}{cc}
             {\bf -0.25}_{[\pm 0.17]} & 0.03_{[\pm 0.24]} \\
            -0.04_{[\pm 0.15]} & -0.03_{[\pm 0.21]} \\
          \end{array}
        \right)$$

$$\hat{A}_5^S=\left(
          \begin{array}{cc}
            -0.06_{[\pm 0.17]} & -0.04_{[\pm 0.19]} \\
            -0.02_{[\pm 0.15]} & -0.05_{[\pm 0.17]} \\
          \end{array}
        \right).$$
We obtain the following $PAM_{OLS}$

$$\hat{A}_1^{OLS}=\left(
          \begin{array}{cc}
            {\bf -0.35}_{[\pm 0.23]} & 0.12_{[\pm 0.23]} \\
            0.06_{[\pm 0.19]} & {\bf -0.72}_{[\pm 0.19]} \\
          \end{array}
        \right)\quad\hat{A}_2^{OLS}=\left(
          \begin{array}{cc}
            {\bf -0.55}_{[\pm 0.27]} & 0.08_{[\pm 0.35]} \\
            0.08_{[\pm 0.24]} & -0.27_{[\pm 0.36]} \\
          \end{array}
        \right)$$

$$\hat{A}_3^{OLS}=\left(
          \begin{array}{cc}
            -0.32_{[\pm 0.31]} & 0.07_{[\pm 0.35]} \\
            0.15_{[\pm 0.18]} & -0.20_{[\pm 0.31]} \\
          \end{array}
        \right)\quad\hat{A}_4^{OLS}=\left(
          \begin{array}{cc}
             -0.25_{[\pm 0.24]} & 0.03_{[\pm 0.26]} \\
            -0.04_{[\pm 0.26]} & -0.03_{[\pm 0.27]} \\
          \end{array}
        \right)$$

$$\hat{A}_5^{OLS}=\left(
          \begin{array}{cc}
            -0.06_{[\pm 0.25]} & -0.04_{[\pm 0.17]} \\
            -0.02_{[\pm 0.19]} & -0.05_{[\pm 0.24]} \\
          \end{array}
        \right).$$and the following $PAM_{ALS}$

$$\hat{A}_1^{ALS}=\left(
          \begin{array}{cc}
          {\bf -0.42}_{[\pm 0.18]} & 0.06_{[\pm 0.24]} \\
            0.02_{[\pm 0.11]} & {\bf -0.70}_{[\pm 0.21]} \\
            \end{array}
        \right)\quad\hat{A}_2^{ALS}=\left(
          \begin{array}{cc}
            {\bf -0.58}_{[\pm 0.19]} & 0.07_{[\pm 0.30]} \\
            0.03_{[\pm 0.12]} & -0.26_{[\pm 0.26]} \\
          \end{array}
        \right)$$

$$\hat{A}_3^{ALS}=\left(
          \begin{array}{cc}
            -0.21_{[\pm 0.21]} & 0.07_{[\pm 0.30]} \\
            0.08_{[\pm 0.13]} & -0.21_{[\pm 0.26]} \\
          \end{array}
        \right)\quad\hat{A}_4^{ALS}=\left(
          \begin{array}{cc}
            -0.20_{[\pm 0.20]} & 0.06_{[\pm 0.31]} \\
            0.02_{[\pm 0.13]} & -0.01_{[\pm 0.27]} \\
          \end{array}
        \right)$$

$$\hat{A}_5^{ALS}=\left(
          \begin{array}{cc}
            -0.13_{[\pm 0.19]} & 0.06_{[\pm 0.26]} \\
            -0.01_{[\pm 0.12]} & -0.07_{[\pm 0.22]} \\
          \end{array}
        \right).$$
It can be seen that the $PAM_{OLS}$ and $PAM_{ALS}$ in the $\hat{A}_i^{OLS}$ and $\hat{A}_i^{ALS}$ for $i=3,4$ and $5$ seem not significant, so that one can identify $p=2$ using our modified tools. The cut-off at $p=2$ is clearly marked for the $PAM_{OLS}$ and $PAM_{ALS}$. If the $PAM_{S}$ are used we note that one could select again $p=3$ or even $p=4$, and we note that the cut-off is not so clearly marked in this case. We also see that the 95\% standard and OLS confidence bounds can be quite different. If the length was automatically selected using the $PAM$ and $PCM$, larger lag lengths would have been chosen. Indeed we note that some of the $PAM$ and $PCM$ are only slightly beyond the 95\% confidence bounds (see for instance the $\hat{P}_{OLS}(3)$, $\hat{P}_{OLS}(4)$ or $\hat{P}_{ALS}(3)$ in Figures \ref{autocorrols1} and \ref{autocorrols2}).\\

In general it emerges from our empirical study part that the
standard identification tools lead to select large lag lengths for
the VAR models with non constant variance. This may be viewed as a
consequence to the fact that the standard tools are not adapted to
our non standard framework. Note that the identification of the
model is the first step of the VAR modeling of time series. In such
situation the practitioner is likely to adjust a VAR model with a
too large number of parameters which can affect the analysis of the
series. The identification tools developed in this paper take into
account for unconditional heteroscedasticity. From our real data
study we found that the modified tools are more parsimonious.

\section*{References}
\begin{description}
\item[]{\sc Ahmed, S., Levin, A., and Wilson, B.A.} (2002) "Recent US macroeconomic stability: Good luck, good policies, or good practices?" \textit{International Finance Discussion Papers}, The board of governors of the federal reserve system, 2002-730.
\item[]{\sc Akaike , H.} (1973) Information theory and an extension of the maximum likelihood principle, in: B.N. Petrov and F. Csaki, eds., \textit{2nd International Symposium on Information Theory}, pp. 267-281. Akademia Kiado, Budapest.
\item[]{\sc Ansley, C.F., and Newbold, P.} (1979) Multivariate partial autocorrelations. \textit{ASA Proceedings of the Business and Economic Statistics Section}, 349-353.
\item[]{\sc Aue, A., H\"{o}rmann, S., Horv\`{a}th, L., and Reimherr,
M.} (2009) Break detection in the covariance structure of
multivariate time series models. \textit{Annals of Statistics} 37,
4046-4087.
\item[] {\sc Bai, J. (2000)} Vector autoregressive models with structural changes in
regression coefficients and in variance-covariance matrices.
\textit{Annals of Economics and Finance} 1, 303-339.
\item[]{\sc Boubacar Mainassara, Y.} (2012) Selection of weak VARMA models by Akaike's information criteria. \textit{Journal of Time Series Analysis} 33, 121-130.
\item[]{\sc Brockwell, P.J., and Davis, R.A.} (1991) \textit{Time Series: Theory and methods}. Springer, New York.

\item[]{\sc Cavanaugh, J.E.} (1997) Unifying the derivations for the Akaike and corrected Akaike information criteria. \textit{Statistics and Probability Letters} 33, 201-208.
\item[]{\sc Dahlhaus, R.} (1997) Fitting time series models to nonstationary processes. \textit{Annals of
Statistics} 25, 1-37.
\item[]{\sc Dahlhaus, R., and Subba Rao, S.} (2006) Statistical inference for time-varying ARCH
processes. \textit{Annals of Statistics} 34, 1075-1114.
\item[]{\sc Davis, S.J. and Kahn, J.A.} (2008) Interpreting the great moderation: Changes in the volatility of economic activity at the macro and micro levels. \textit{Journal of Economic Perspectives} 22, 155-180.
\item[]{\sc Engle, R.F., and Rangel, J.G.} (2008) The spline GARCH model for low-frequency volatility and its global macroeconomic causes. Working paper, Czech National Bank.
\item[]{\sc  Francq, C., and Gautier, A.} (2004) Large sample properties of parameter least squares estimates for
time-varying ARMA models. \textit{Journal of Time Series Analysis}
25, 765-783.
\item[]{\sc Galeano, P., and Pena, D.} (2007) Covariance changes detection in multivariate time series. \textit{Journal of Statistical Planning and Inference} 137, 194-211.
\item[]{\sc Hannan, E.J., and Quinn, B.G.} (1979) The determination of the order of an autoregression. \textit{Journal of the Royal Statistical Society B} 41, 190-195.
\item[]{\sc Herrera, A.M., and Pesavento, E.} (2005) The decline in the US output volatility: Structural changes and inventory investment. \textit{Journal of Business and Economic Statistics} 23, 462-472.
\item[]{\sc Horv\`{a}th, L., Kokoszka, P. and Zhang, A.} (2006) Monitoring constancy of variance in conditionally heteroskedastic time series. \textit{Econometric Theory} 22, 373-402.
\item[]{\sc Horv\`{a}th, L., and Steinebach, J.} (2000) Testing for changes in the mean or variance of a stochastic
process under weak invariance. \textit{Journal of Statistical
Planning and Inference} 91, 365-376.
\item[]{\sc Hurvich, C.M., and Tsai, C.-L.} (1989) Regression and time series model selection in small samples. \textit{Biometrika} 76, 297-307.
\item[]{\sc Kim, C.S., and Park, J.Y.} (2010) Cointegrating regressions with time heterogeneity.
\textit{Econometric Reviews} 29, 397-438.
\item[]{\sc Kim, C.J., and Nelson, C.R.} (1999) Has the U.S. economy become more stable? A bayesian approach based on a Markov-switching model of the business cycle. \textit{The Review of Economics and Statistics} 81, 608-616.
\item[] {\sc Kokoszka, P., and Leipus, R.} (2000) Change-point estimation in ARCH models. \textit{Bernoulli} 6, 513-539.
\item[]{\sc Lütkepohl, H.} (2005) \!\textit{New Introduction to Multiple Time Series Analysis}. Springer, Berlin.
\item[]{\sc McConnell, M.M., Mosser P.C. and Perez-Quiros, G.} (1999) A decomposition of the increased stability of GDP growth. \textit{Current Issues in Economics and Finance}, Federal Reserve Bank of New York.
\item[]{\sc Mikosch, T., and St\u{a}ric\u{a}, C.} (2004) Nonstationarities in financial time series, the long-
range dependence, and the IGARCH effects. \textit{Review of
Economics and Statistics} 86, 378-390.
\item[]{\sc Patilea, V., and Raïssi, H.} (2012) Adaptive estimation of vector
autoregressive models with time-varying variance: application to
testing linear causality in mean.
\textit{Journal of Statistical Planning and Inference} 142, 2891-2912.   
\item[]{\sc Patilea, V., and Raïssi, H.} (2011) Corrected portmanteau tests for VAR models with time-varying variance. Working paper, Université européenne de Bretagne IRMAR-INSA.
\item[]{\sc Paulsen, J.} (1984) Order determination of multivariate autoregressive time series with unit roots, \textit{Journal of Time Series Analysis} 5, 115-127.
\item[]{\sc Pesaran, H., and Timmerman, A.} (2004) How costly is it to ignore breaks when
forecasting the direction of a time series. \textit{International
Journal of Forecasting} 20, 411-425.
\item[]{\sc Qu, Z., and Perron, P.} (2007) Estimating and testing structural changes in multivariate regressions. \textit{Econometrica} 75, 459-502.
\item[]{\sc Reinsel, G.C.} (1993) \textit{Elements of Multivariate Time Series Analysis}. Springer, New-York.
\item[]{\sc Sanso, A., Arag\'{o}, V., and Carrion, J.L. } (2004). Testing for changes in the unconditional variance of financial time series. \textit{Revista de Economia Financiera} 4, 32-53.
\item[]{\sc Sensier, M., and van Dijk, D.} (2004) Testing for volatility changes in U.S. macroeconomic time series. \textit{Review of Economics and Statistics} 86, 833-839.
\item[]{\sc St\u{a}ric\u{a}, C.} (2003) Is GARCH(1,1) as good a model as the Nobel
prize accolades would imply?. Working paper,
http://129.3.20.41/eps/em/papers/0411/0411015.pdf.
\item[]{\sc Stock, J.H., and Watson, M.W.} (2002) "Has the business cycle changed and why?", NBER Macroannual 2002, M. Gertler and K. Rogoff eds., MIT Press.
\item[]{\sc Tiao, G.C., and Box, G.E.P.} (1981) Modeling multiple times series with applications. \textit{Journal of the American Statistical Association} 76, 802-816.
\item[]{\sc Xu, K.L., and Phillips, P.C.B.} (2008)
Adaptive estimation of autoregressive models with time-varying
variances. \textit{Journal of Econometrics} 142, 265-280.
\end{description}

\newpage
\section*{Appendix: Tables and Figures}

\begin{table}[hh]\!\!\!\!\!\!\!\!\!\!
\begin{center}
\caption{\small{Frequency (in \%) of selected lag length.
The simulated processes are of length $n=50$ with variance specified
as in (\ref{smooth}).}}
\vspace*{0.1cm}
\begin{tabular}{|c|c|c|c|c|c|}
\hline
  $p$ & 1 &{\bf 2} & 3 & 4 & 5  \\
  \hline
  $AIC$ & 1.5& 0.2 & 0.1& 0.1& {\bf 98.1}\\
  \hline
  $AIC_{ALS}$ &8.0  &{\bf 58.4}  &15.6  &9.0 & 9.0 \\
  \hline
  $AIC_{GLS}$ & 7.2& {\bf 84.0} & 6.4 & 1.8& 0.6 \\
  \hline
\end{tabular}
\label{tab1}
\end{center}
\end{table}

\begin{table}[hh]\!\!\!\!\!\!\!\!\!\!
\begin{center}
\caption{\small{The same as in Table \ref{tab1} but for $n=100$.}}
\vspace*{0.1cm}
\begin{tabular}{|c|c|c|c|c|c|}
\hline
  $p$ & 1 &{\bf 2} & 3 & 4 & 5  \\
  \hline
  $AIC$ & 0.0& 0.0 & 0.0& 0.0& {\bf 100.0}\\
  \hline
  $AIC_{ALS}$ &2.4  &{\bf 78.1}  &11.9  &5.5 & 2.1 \\
  \hline
  $AIC_{GLS}$ & 0.2& {\bf 86.6} & 9.8 & 2.4& 1.0\\
  \hline
\end{tabular}
\label{tab2}
\end{center}
\end{table}

\begin{table}[hh]\!\!\!\!\!\!\!\!\!\!
\begin{center}
\caption{\small{The same as in Table \ref{tab1} but for $n=200$.}}
\vspace*{0.1cm}
\begin{tabular}{|c|c|c|c|c|c|}
\hline
  $p$ & 1 &{\bf 2} & 3 & 4 & 5 \\
  \hline
  $AIC$ & 0.0& 0.0 & 0.0& 0.0& {\bf 100.0}\\
  \hline
  $AIC_{ALS}$ &1.6  &{\bf 84.1}  &9.3  &3.8 & 1.2\\
  \hline
  $AIC_{GLS}$ & 0.0& {\bf 89.1} & 6.6 & 3.8& 0.5\\
  \hline
\end{tabular}
\label{tab3}
\end{center}
\end{table}


\begin{table}[hh]\!\!\!\!\!\!\!\!\!\!
\begin{center}
\caption{\small{Frequency (in \%) of selected lag length.
The simulated processes are of length $n=50$ with variance specified
as in (\ref{break}).}}
\vspace*{0.1cm}
\begin{tabular}{|c|c|c|c|c|c|}
\hline
  $p$ & 1 &{\bf 2} & 3 & 4 & 5  \\
  \hline
  $AIC$ & 3.3& 0.2 & 0.4& 0.7& {\bf 95.4}\\
  \hline
  $AIC_{ALS}$ &19.2  &{\bf 44.7}  &16.6  &11.3 & 8.2 \\
  \hline
  $AIC_{GLS}$ & 7.5& {\bf 79.8} & 8.3 & 3.0& 1.4 \\
  \hline
\end{tabular}
\label{tab1b}
\end{center}
\end{table}

\begin{table}[hh]\!\!\!\!\!\!\!\!\!\!
\begin{center}
\caption{\small{The same as in Table \ref{tab1b} but for $n=100$.}}
\vspace*{0.1cm}
\begin{tabular}{|c|c|c|c|c|c|}
\hline
  $p$ & 1 &{\bf 2} & 3 & 4 & 5  \\
  \hline
  $AIC$ & 0.4& 0.1 & 0.0& 0.0& {\bf 99.5}\\
  \hline
  $AIC_{ALS}$ &19.3  &{\bf 63.7}  &11.2  &4.3 & 1.5\\
  \hline
  $AIC_{GLS}$ & 0.3& {\bf 84.6} & 9.9 & 3.5& 1.7 \\
  \hline
\end{tabular}
\label{tab2b}
\end{center}
\end{table}

\begin{table}[hh]\!\!\!\!\!\!\!\!\!\!
\begin{center}
\caption{\small{The same as in Table \ref{tab1b} but for $n=200$.}}
\vspace*{0.1cm}
\begin{tabular}{|c|c|c|c|c|c|}
\hline
  $p$ & 1 &{\bf 2} & 3 & 4 & 5 \\
  \hline
  $AIC$ & 0.0& 0.0 & 0.0& 0.0& {\bf 100.0}\\
  \hline
  $AIC_{ALS}$ &3.9  &{\bf 77.3}  &11.1  &5.6 & 2.1\\
  \hline
  $AIC_{GLS}$ & 0.0& {\bf 86.5} & 8.2 & 4.3& 1.0 \\
  \hline
\end{tabular}
\label{tab3b}
\end{center}
\end{table}

\begin{table}[hh]\!\!\!\!\!\!\!\!\!\!
\begin{center}
\caption{\small{Frequency (in \%) of PAM and PCM for parameters
$A_{0p}(1,1)$ beyond of their 95\% asymptotic confidence bounds.
The simulated processes are of length $n=50$ with variance specified
as in (\ref{smooth}).}}
\vspace*{0.1cm}
\begin{tabular}{|c|c|c|c|c|c|}
\hline
  $p$ & 1 &{\bf 2} & 3 & 4 & 5  \\
  \hline
  $PAM_S$ & 72.4& 93.6 & 11.4& 11.3& 12.8\\
  \hline
  $PAM_{OLS}$ &69.0  &91.9  &9.6  &10.7 & 11.6 \\
  \hline
  $PAM_{ALS}$ &69.8  &92.8  &10.1  &9.9 & 10.7 \\
  \hline
  $PAM_{GLS}$ & 64.3& 89.9 & 5.1 & 5.1& 4.6 \\
  \hline  \hline
  $PCM_S$ & 39.1& 94.0 & 10.6& 10.0& 12.6\\
  \hline
  $PCM_{OLS}$ &35.9  & 92.6  &9.2  &10.2 & 11.8 \\
  \hline
  $PCM_{ALS}$ &34.4  &94.1  &8.2  &8.9 & 11.2 \\
  \hline
  $PCM_{GLS}$ & 55.5& 94.3 & 4.4 & 4.7& 5.3 \\
  \hline
\end{tabular}
\label{tab1pampcm}
\end{center}
\end{table}

\begin{table}[hh]\!\!\!\!\!\!\!\!\!\!
\begin{center}
\caption{\small{The same as in Table \ref{tab1} but for $n=100$.}}
\vspace*{0.1cm}
\begin{tabular}{|c|c|c|c|c|c|}
\hline
  $p$ & 1 & {\bf 2} & 3 & 4 & 5  \\
  \hline
  $PAM_S$ & 94.5& 100.0 & 9.3& 9.2& 7.6\\
  \hline
  $PAM_{OLS}$ &93.3  &100.0  &7.5  &7.5 & 6.0 \\
  \hline
  $PAM_{ALS}$ &94.6  &100.0  &7.6  &7.7 & 5.0 \\
  \hline
  $PAM_{GLS}$ & 93.4& 99.9 & 4.8 & 5.6& 3.6 \\
  \hline  \hline
  $PCM_S$ & 81.7&  99.9 & 8.7& 9.1&8.1\\
  \hline
  $PCM_{OLS}$ &74.4  &100.0  &6.7  &7.4 & 6.6 \\
  \hline
  $PCM_{ALS}$ &79.5  &100.0  &6.7  &6.5 & 5.6 \\
  \hline
  $PCM_{GLS}$ &93.6& 100.0 & 4.8 & 5.2& 4.2  \\
  \hline
\end{tabular}
\label{tab2pampcm}
\end{center}
\end{table}

\begin{table}[hh]\!\!\!\!\!\!\!\!\!\!
\begin{center}
\caption{\small{The same as in Table \ref{tab1} but for $n=200$.}}
\vspace*{0.1cm}
\begin{tabular}{|c|c|c|c|c|c|}
\hline
  $p$ & 1 &{\bf 2} & 3 & 4 & 5 \\
  \hline
  $PAM_S$ & 99.9& 100.0 & 10.1& 10.1& 8.7\\
  \hline
  $PAM_{OLS}$ &99.8  &100.0  &6.2  &7.7 & 5.6 \\
  \hline
  $PAM_{ALS}$ &100.0  &100.0  &6.3  &6.3 & 5.8 \\
  \hline
  $PAM_{GLS}$ & 99.9& 100.0 & 6.1 & 5.0& 5.4 \\
  \hline  \hline
  $PCM_S$ & 99.4& 100.0 & 8.1& 8.5& 8.5\\
  \hline
   $PCM_{OLS}$ &98.4  &100.0  &5.4  &6.0 & 6.1 \\
  \hline
  $PCM_{ALS}$ &99.6  &100.0  &5.2  &6.6 & 6.2 \\
  \hline
  $PCM_{GLS}$ & 100.0& 100.0 & 4.8 & 5.6& 5.9  \\
  \hline
\end{tabular}
\label{tab3pampcm}
\end{center}
\end{table}


\begin{table}[hh]\!\!\!\!\!\!\!\!\!\!
\begin{center}
\caption{\small{Frequency (in \%) of PAM and PCM for parameters
$A_{0k}(1,1)$ beyond of their 95\% asymptotic confidence bounds.
The simulated processes are of length $n=50$ with variance specified
as in (\ref{break}).}}
\vspace*{0.1cm}
\begin{tabular}{|c|c|c|c|c|c|}
\hline
  $p$ & 1 &{\bf 2} & 3 & 4 & 5  \\
  \hline
  $PAM_S$ & 66.4& 89.2 & 16.9& 18.4& 19.0\\
  \hline
  $PAM_{OLS}$ &58.4 & 84.1  &10.8  &15.4 &14.3 \\
  \hline
  $PAM_{ALS}$ &63.0  &88.6  &11.4  &11.1 & 12.3 \\
  \hline
  $PAM_{GLS}$ & 56.9& 85.9 & 15.1 & 8.5& 7.4 \\
  \hline  \hline
  $PCM_S$ & 38.2& 90.3 & 17.1& 17.3& 19.9\\
  \hline
  $PCM_{OLS}$ &22.9  &85.5  &10.7  &13.5 & 13.6 \\
  \hline
  $PCM_{ALS}$ &27.4  &89.3  &10.0  &10.4 & 13.0\\
  \hline
  $PCM_{GLS}$ & 51.6& 92.9 & 13.4 & 10.4& 6.9 \\
  \hline
\end{tabular}
\label{tab1bpampcm}
\end{center}
\end{table}

\begin{table}[hh]\!\!\!\!\!\!\!\!\!\!
\begin{center}
\caption{\small{The same as in Table \ref{tab1bpampcm} but for
$n=100$.}}
\vspace*{0.1cm}
\begin{tabular}{|c|c|c|c|c|c|}
\hline
  $p$ & 1 &{\bf 2} & 3 & 4 & 5  \\
  \hline
  $PAM_S$ & 91.4& 99.6 & 16.3& 16.1& 15.5\\
  \hline
  $PAM_{OLS}$ &83.9  &98.5  &8.7  &9.5 & 8.2 \\
  \hline
  $PAM_{ALS}$ &89.7  &99.6  &7.9  &8.7 & 6.9 \\
  \hline
  $PAM_{GLS}$ & 90.6& 99.8 & 11.8 & 8.8& 6.0 \\
  \hline  \hline
  $PCM_S$ & 76.7& 99.7 & 14.5& 15.5& 14.2\\
  \hline
  $PCM_{OLS}$ &52.3  &99.1  &8.0  &8.6 & 8.6 \\
  \hline
  $PCM_{ALS}$ &66.1  &99.8  &7.9  &7.9 & 7.2 \\
  \hline
  $PCM_{GLS}$ & 92.4& 100.0 & 10.4 & 9.4& 6.2 \\
  \hline
\end{tabular}
\label{tab2bpampcm}
\end{center}
\end{table}

\begin{table}[hh]\!\!\!\!\!\!\!\!\!\!
\begin{center}
\caption{\small{The same as in Table \ref{tab1bpampcm} but for
$n=200$.}}
\vspace*{0.1cm}
\begin{tabular}{|c|c|c|c|c|c|}
\hline
  $p$ & 1 & {\bf 2} & 3 & 4 & 5 \\
  \hline
  $PAM_S$ & 99.4& 100.0 & 16.5& 16.3& 16.1\\
  \hline
  $PAM_{OLS}$ &97.9  &100.0  &6.1  &7.1 & 5.9 \\
  \hline
  $PAM_{ALS}$ &99.9  &100.0  &8.8  &7.7 & 5.9  \\
  \hline
  $PAM_{GLS}$ & 99.8& 100.0 & 10.9 & 9.5&7.3 \\
  \hline  \hline
  $PCM_S$ & 98.2& 100.0 & 14.9& 14.8& 15.4\\
  \hline
  $PCM_{OLS}$ &90.2  &100.0  &5.8  &7.2 & 5.9 \\
  \hline
  $PCM_{ALS}$ &98.6  &100.0  &6.9  &6.6 & 6.3\\
  \hline
  $PCM_{GLS}$ & 100.0& 100.0 & 8.4 & 7.8& 7.4 \\
  \hline
\end{tabular}
\label{tab3bpampcm}
\end{center}
\end{table}

\begin{table}[hh]\!\!\!\!\!\!\!\!\!\!
\begin{center}
\caption{\small{The quarterly foreign direct investment and
government securities hold by foreigners for the U.S. in billions dollars ($n=147$): The
selected autoregressive order using $AIC$ and $AIC_{ALS}$.}}
\vspace*{0.1cm}
\begin{tabular}{|c|c|c|c|c|c|}
\hline
  $p$ & 1 & 2 & 3 & 4 & 5  \\
  \hline
  $AIC$ &13.91 & 13.73 & 13.59 & 13.53 & {\bf 13.52} \\
  \hline
  $AIC_{ALS}$ &11.80 & {\bf 11.61} & 11.74 & 11.77 & 11.80 \\
  \hline 
\end{tabular}
\label{realdata}
\end{center}
\end{table}

%
%
%
%
%

\end{document}